\documentclass[prb, showpacs, floats, superscriptaddress, twocolumn]{revtex4-1}
\usepackage[utf8x]{inputenc}
\usepackage{amsfonts}
\usepackage[english]{babel} 
\usepackage[T1]{fontenc}
\usepackage{amsmath}
\usepackage{bbold}
\usepackage{graphicx}
\usepackage{hyperref}   % use for hypertext links, including those to external documents and URLs
\usepackage{color}
\usepackage{url}
\usepackage{natbib}
\usepackage{hyperref}
\hypersetup{
 	colorlinks=true,
	linkcolor=blue,
	citecolor=blue
}

\DeclareMathOperator{\Tr}{Tr}

\newcommand{\pp}{\text{:}}

\newcommand{\phin}{{\phi}^{\eta}} 
\newcommand{\phinp}{{\phi}^{\eta'}} 
\newcommand{\phinpp}{{\phi}^{\eta''}} 
\newcommand{\phinppp}{{\phi}^{\eta'''}}

\newcommand{\intinf}{{\int}_{-\infty}^{\infty}}

\newcommand{\ptr}[1]{{\partial_{t_{#1}}} }
 
\newcommand{\ellll}{C^{\lambda_2 \lambda_3 \lambda_4 \lambda_5}_{\eta, \eta', \eta'', \eta'''}(2,3,4,5)} 
\newcommand{\eppmm}{C^{+ + - -}_{\eta, \eta', \eta'', \eta'''}(2,3,4,5)} 
\newcommand{\emmpp}{C^{- - + +}_{\eta, \eta', \eta'', \eta'''}(2,3,4,5)} 
\newcommand{\epmpm}{C^{+ - + -}_{\eta, \eta', \eta'', \eta'''}(2,3,4,5)} 
 
\newcommand{\epmmp}{C^{+ - - +}_{\eta, \eta', \eta'', \eta'''}(2,3,4,5)}

\newcommand{\cnull}{C^{\eta, \eta', \eta'', \eta'''}_0(t_2,t_3,t_4,t_5)} 
 
\newcommand{\ceinsp}{C_1^{+}(t_2,t_3,t_4,t_5)} 
\newcommand{\ceinsm}{C_1^{-}(t_2,t_3,t_4,t_5)} 
 
\newcommand{\ceinspm}{C_1^{\pm}(t_2,t_3,t_4,t_5)} 

\newcommand{\Hnull}{H_{0}} 
\newcommand{\Hll}{H_{LL}} 
\newcommand{\Hint}{H_{\textrm{int}}} 
\newcommand{\hV}{H_V}
\newcommand{\hVt}{{H'}_V}
\newcommand{\Hr}{{H}_R} 
\newcommand{\Hm}{{H}_m} 
\newcommand{\Hc}{\textrm{H.c.}} 
\newcommand{\jtr}{j_{\textrm{tr}} }
\newcommand{\jfr}{j_{0} }
\newcommand{\jbs}{j_{\textrm{bs}} }

\newcommand{\Ud}{U^{\dagger}}
\newcommand{\Ht}{{\tilde{H}}_{R}}
\newcommand{\Hti}{\left({\tilde{H}}_{R}\right)_I}

\newcommand{\Oti}{\left({\tilde{\mathcal{O}}}\right)_I}

\newcommand{\jfrti}{\left(\tilde{j}_{0}\right)_I}

\newcommand{\jbsti}{\left(\tilde{j}_{\textrm{bs}}\right)_I}

\newcommand{\Psid}{\Psi^{\dagger}}

\newcommand{\Ps}[2]{\Psi_{#1,#2} }
\newcommand{\Psd}[2]{\Psi^{\dagger}_{#1,#2}}
\newcommand{\PsS}[1]{\Psi_{#1} }
\newcommand{\PsdS}[1]{\Psi^{\dagger}_{#1}}

\begin{document}

% \title{Non-equilibrium transport at the interacting quantum spin Hall edge with a local Rashba spin-orbit scatterer}
%  \title{Single- and two-particle backscattering off a local Rashba spin-orbit scatterer at the quantum spin Hall edge}
\title{Evidence for broken Galilean invariance at the quantum spin Hall edge}
\author{Florian Geissler}
  \affiliation{Institute for Theoretical Physics and Astrophysics,
 University of W\"urzburg, 97074 W\"urzburg, Germany}
\author{Fran\c{c}ois Crépin}
  \affiliation{Institute for Theoretical Physics and Astrophysics,
 University of W\"urzburg, 97074 W\"urzburg, Germany}
\author{Bj\"orn Trauzettel}
  \affiliation{Institute for Theoretical Physics and Astrophysics,
 University of W\"urzburg, 97074 W\"urzburg, Germany}
%  \author{Thomas Schmidt}
%   \affiliation{Institute for Theoretical Physics and Astrophysics,
%  University of W\"urzburg, 97074 W\"urzburg, Germany}

\date{\today}

 \begin{abstract}
 
 We study transport properties of the helical edge channels of a quantum spin Hall (QSH) insulator, in the presence of electron-electron interactions and weak, local Rashba spin-orbit coupling. 
 The combination of the two allows for inelastic backscattering that does not break time-reversal symmetry (TRS), resulting in interaction-dependent power law corrections to the conductance.
 Here, we use a non-equilibrium Keldysh formalism to describe the situation of a long, one-dimensional edge channel coupled to external reservoirs, where the applied bias is the leading energy scale.
 By calculating explicitly the corrections to the conductance up to fourth order of the impurity strength, we analyse correlated single- and two-particle backscattering processes on a microscopic level.
 Interestingly, we show that the modeling of the leads together with the breaking of Galilean invariance has important consequences on the transport properties.
 Such breaking occurs, because the Galilean invariance of the bulk spectrum transforms into an emergent Lorentz invariance of the edge spectrum.
 With this broken Galilean invariance at the QSH edge, we find a contribution to single particle backscattering with a very low power scaling, while in the presence of Galilean invariance 
 the leading contribution would be due to correlated two-particle backscattering only. This difference is further reflected in different values of the Fano factor of the shot noise, an experimentally observable quantity.
 The described behaviour is specific to the Rashba scatterer, and does not occur in the case of backscattering off a time-reversal breaking, magnetic impurity.
 \end{abstract}

 \maketitle
 
\section{Introduction}

Quantum spin Hall systems \cite{KaneMele, KaneMele2, Ber06} are prominent examples of a topologically non-trivial phase of matter. Even though the bulk system is gapped, gapless edge
states with a linear energy dispersion are formed at the boundaries of the (two-dimensional) QSH sample, in contact with vacuum or a normal insulator. These one-dimensional edge channels are responsible for peculiar transport properties and
contribute to the conductance with $G_0=e^2/h$ per channel. Such quantized values of the conductance have been experimentally observed e.g. in HgTe/CdTe~\cite{Mole} or InAs/GaSb quantum wells~\cite{Kne11}.\\
The ballistic edge transport in the QSH system is protected by two mechanisms - the helical character of the transport channels and time-reversal symmetry (TRS). 
First, helicity is an essential feature of the QSH edge states. The electrons direction of motion is 
strongly coupled to the spin direction, meaning that electrons propagating in opposite directions are expected to have opposite spins. In inversion-preserving geometries, we can describe the edge electrons by the two
orthogonal eigenstates of the $s_z$ spin operator. 
% $(\psi_{\uparrow,k},\psi_{\downarrow,-k})$. 
With spin-orbit coupling (SOC) present, the spin $s_z$ component is not 
preserved any more, however, Kramers theorem ensures the existence of a pair of spin-orthogonal states 
% $(\psi_{+,k},\psi_{-,-k})$ 
with opposite momentum, 
such that the helical character remains \cite{Sch12}. By this virtue, the helical edge transport is insensitive to disorder that does not flip the electrons spin, since there is simply no 
adequate propagation channel for the backscattered electron. In the presence of spin-flipping impurities, time-reversal-symmetry still prevents elastic backscattering between counter-propagating channels \cite{KaneMele,Xu06,Wu06}.\\ 
Importantly however, these mechanisms do not protect the helical edge against inelastic backscattering.
% (nor against backscattering off a TRS-breaking impurity \cite{MaciejkoOreg09}).
% In particular, a TRS-compatible combination of spin-orbit coupling and an inelastic scattering channel provided by 
Coulomb interactions or phonons will lead to single- (SPB) or multiple-particle backscattering \cite{Stroem, Crepin1, Sch12, Budich, Oreg}. Especially,
the two-particle backscattering (TPB) 
processes can be a relevant perturbation for strong interaction strengths, and open up a gap in the spectrum, while SPB contributions were previously shown to be vanishing \cite{Budich}. Such findings have been achieved essentially with the help of renormalization group (RG) approaches, using 
finite temperatures at equilibrium \cite{Stroem, Crepin1, GeissCrep14}.
In Ref.~\onlinecite{Oreg}, a non-equilibrium framework was used to study the correction to the conductance at finite voltage bias due to generic, inelastic backscattering.\\
In this work, we study both single and two-particle backscattering off a single, local Rashba scatterer,
starting from a microscopic picture. To account for a more realistic situation, we apply a voltage bias on two non-interacting leads connected to 
an interacting helical Luttinger liquid and use a non-equilibrium Keldysh formalism to calculate the expected backscattering current up to fourth order of the impurity strength. 
Somewhat surprisingly, the modeling of the external reservoirs as well as the rigorous treatment of elementary invariances, such as the Pauli principle and Galilean invariance, have qualitative effects on the resulting current.
In particular, we discover a not yet reported single-particle contribution, that may arise in the absence of Galilean invariance and the aforementioned design of the leads. 
The corresponding correction to the conductance is found to scale with a remarkably low power of $\delta G \sim V^2$ in the limit of weak interactions.
On the other hand, if Galilean invariance were present, the lowest order contribution would stem from correlated two-particle processes. Besides the different power law 
scalings of the correction to the conductance, the distinct character of the two situations is also reflected in the Fano factor, as we verify calculating the shot noise. In the 
case of broken Galilean invariance, we predict a Fano factor of one, while in the presence of Galilean symmetry, correlated two-particle 
backscattering processes become manifest in a Fano factor of two.
We therefore present an experimentally accessible quantity, the shot noise, as a direct evidence of broken Galilean invariance in this setup.
For a better understanding of the problem, we furthermore work out the differences with the case of a TRS-breaking impurity.\\
The article is organized as follows: In Sec. II, we give a detailed description of the model, deriving the effective operators in the presence of the external bias. While electron-electron 
interactions are treated exactly in the framework of bosonization, we perform a perturbative expansion in the Rashba impurity strength. In Secs. III and IV, the averaged backscattering current, 
as well as the shot noise, are calculated perturbatively in second and fourth order of the impurity strength, respectively. 
In order to verify our findings in Sec. III.A, an alternative calculation in fermionic language is presented in Sec. III.B.  
Finally, in Sec. V, we summarize our results. Additional details are given in the Appendices.

\section{Model}

\subsection{Bosonization}
\begin{figure}[htb]
\includegraphics[width=0.5 \textwidth]{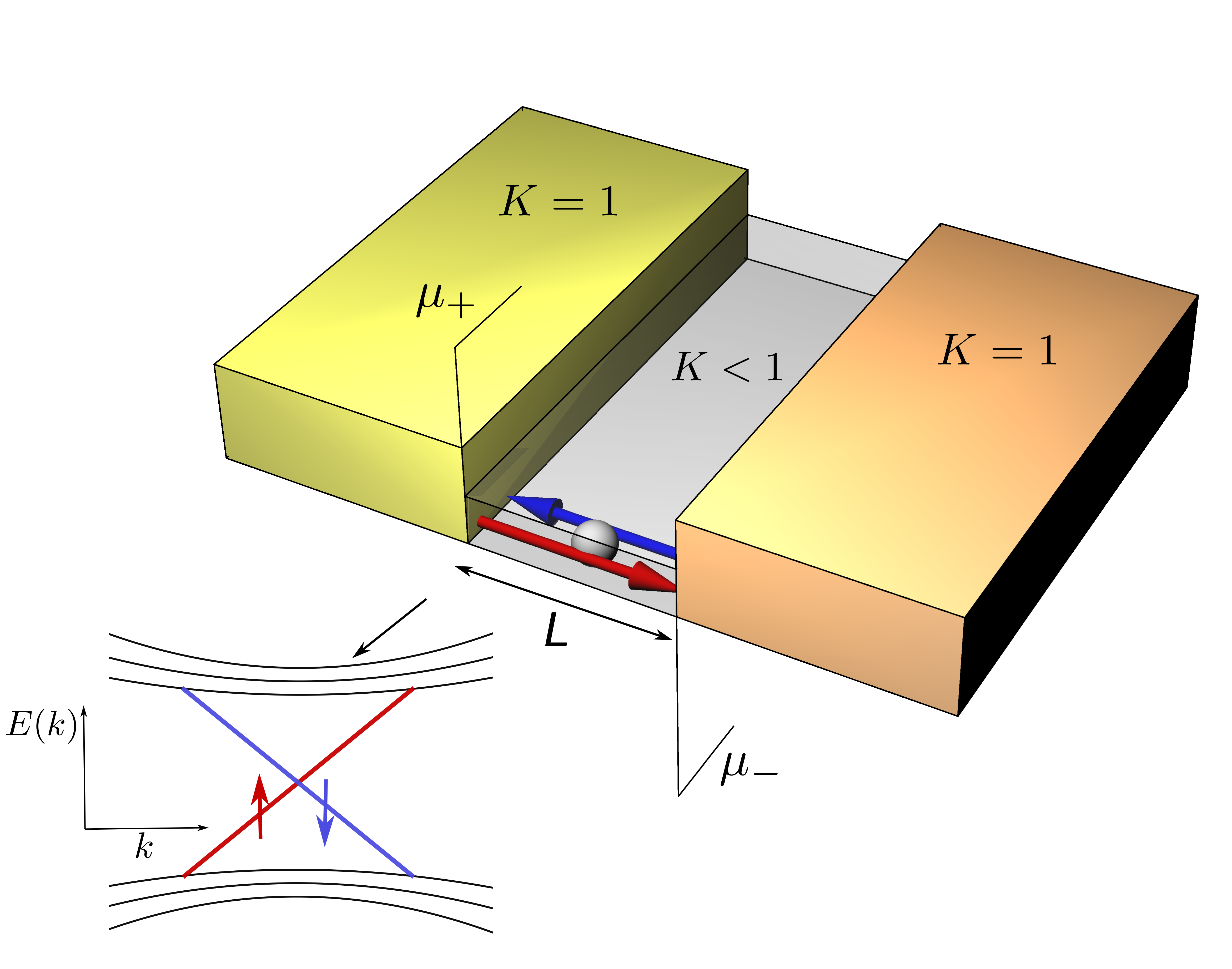}
\caption{(Color online) Scheme of the interacting QSH edge channel, attached to two long, non-interacting leads. 
% We assume the system length $L$ to be large compared to the other lengthscales of the system $L\gg 1/eV \gg a$.
The external bias modifies the number of right and left moving electrons in the contacts. Transport along the edge channel is interrupted by a local Rashba SOC scatterer (dot). The inset 
shows the linear energy spectrum of the edge state electrons.}
\label{fig:PlotLeads}
\end{figure}

The system under consideration can be described by a Hamiltonian consisting of the four terms $H=\Hnull+\Hr+\hV+ \Hint$, with
\begin{align}
& \Hnull= \int dx\ \sum_{r=\pm}\Psi^{\dagger}_r(x)\left(-i r v_F \partial_x-\mu_r \right)\Psi_r(x), \notag \\
% & \Hint= g_2 \int dx\ \Psi^{\dagger}_+(x) \Psi^{\dagger}_-(x) \Psi_-(x)\Psi_+(x), \notag \\
& \Hr=\int dx\ \alpha(x)\left[ \left(\partial_x \Psi^{\dagger}_+\right) \Psi_- - \Psi^{\dagger}_+ \left(\partial_x \Psi^{\phantom{\dagger}}_- \right) \right](x)+ \Hc \notag, \\
& \hV= -\frac{1}{\pi} \int dx \left(\mu_+ \Psid_+(x) \Psi_+(x) +\mu_- \Psid_-(x) \Psi_-(x)\right), \notag \\
& \Hint= g_2 \int dx\ \Psi^{\dagger}_+(x) \Psi^{\dagger}_-(x) \Psi_-(x)\Psi_+(x). \notag
\end{align}
$\Psi^{\dagger}_{\pm}(x)$ and $\Psi_{\pm}(x)$ are fermionic creation and annihilation operators for a right $(+)$ or left $(-)$ moving particle 
and $v_F$ the Fermi velocity. 
% $\omega_0$ is a constant to be specified and 
$\mu_+$ and $\mu_-$ are the chemical potentials of the right and left moving particles, respectively. We set $\hbar=1$ in this article unless explicitly stated.
$\Hnull$ describes the free Hamiltonian with a strictly linear dispersion relation \cite{Gia} and $\Hint$ embodies electron-electron interactions, where we take into account only processes of the type $g_2$ coupling 
right and left movers. It is a particular feature of the helical liquid that in the case of contact interactions the terms $g_1$ and $g_4$ of the usual g-ology are absent, as we explain in the Appendix \ref{app:interactions}.
The system is interspersed by perturbations that we choose to be of the form of a weak, Rashba-like impurity $\Hr$ coupling right and left movers \cite{Crepin1,Stroem,GeissCrep14}.
$\alpha(x)$ may be any function that models the presence of Rashba impurities, however, in this article, we will restrict ourselves to a Dirac-like $\delta$-function for simplicity. This impurity Hamiltonian $H_R$ constitutes, from our perspective, the simplest model that can couple the two edge states and preserve TRS at the same time. Note that it has also been derived on the basis of rather general assumptions in the context of the so-called generic helical liquid \cite{Sch12, ImamGlaz12}. \\
%
%
%-----------Bosonization
%
In the following, we will use the technique of bosonization to treat interactions exactly, while including the impurity strength perturbatively. We make use of the bosonization identity~\cite{Haldane, Haldane2, Delft, Gia}
\begin{equation*}
 \Psi_{\pm}(x)=F_{\pm} \frac{1}{\sqrt{2\pi a}}e^{\pm i \frac{2\pi}{L}N_{\pm} x}e^{-i\left(\pm \phi(x) -\theta(x) \right)}\;,
\end{equation*}
where $F_{\pm}$ is the Klein factor for a right/left moving particle, $N_{\pm}$ the particle number operator counting right/left moving particles, $a$ a short-distance cutoff,
% $k_F$ the Fermi momentum, 
and $\phi$ and $\theta$ two bosonic fields obeying the commutation relation\cite{Delft} 
$[\phi(x),\partial_{x'}\theta(x')]=i \pi (\delta(x-x')-\frac{1}{L})$. In the limit of a large wire length, $L\to \infty$, the expressions will simplify, however, this 
limit has to be taken with care. The fields $\phi$ and $\theta$ are related to the density and current operators respectively, through
\begin{align*}
:\rho_+(x): + :\rho_-(x): = \frac{1}{L}(N_+ + N_-) - \frac{1}{\pi} \partial_x \phi(x) \;, \\
:\rho_+(x): - :\rho_-(x): = \frac{1}{L}(N_+ - N_-) + \frac{1}{\pi} \partial_x \theta(x) \;. 
\end{align*}
Here, we adopt the definition of the bosonic fields of Ref.~\onlinecite{Gia}, except for that the zero-modes $N_+$ and $N_-$ are written explicitly and are not included in the fields $\phi$ and $\theta$ anymore. We then specify the chemical potential introducing the voltage bias $V$ by $\mu_{\pm}=\pm eV/2$.
The Hamiltonian, in its bosonized form, is now given by $H = \Hll + \hV +\Hr$, with \cite{Crepin1}
\begin{align}
\label{eq:Hll}
 \Hll & =\frac{\pi v_F}{L}(N_{+}^2+N_{-}^2)  \notag \\
  & + \frac{v}{2\pi} \int dx \left[ K \pp(\partial_x \theta(x))^2\pp+ \frac{1}{K}\pp(\partial_x \phi(x))^2\pp \right], \\
\label{eq:hV}
 \hV &=  \frac{eV}{2} (N_+-N_-), \\
\label{eq:HR}
  \Hr & =i F_{+}^{\dagger} F_{-} \int dx \, \frac{\alpha(x)}{\pi a} \left(\frac{2\pi a}{L}\right)^K \notag \\
& \times \pp \left(\partial_x \theta(x) +\frac{\pi}{L}(N_{+}-N_{-}) \right) e^{2i (\phi(x)-\frac{\pi x}{L}(N_{+}+N_{-})) }  \pp  \notag \\
& + \Hc
\end{align}
Operators between colons $\pp (\ldots) \pp$ are normal-ordered.
The free Luttinger liquid Hamiltonian $\Hll$ comprises both the linear dispersion and the Coulomb interactions, embodied in the new sound velocity\cite{Gia} $v= v_F (1- (\tfrac{g_2}{2\pi v_F})^2)^{1/2}$ and the dimensionless
parameter $K= ( \tfrac{1-g_2/(2\pi v_F)}{1+g_2/(2\pi v_F)} )^{1/2}$ . The non-interacting case $g_2\to0$ corresponds to $K\to 1$, while $0<K<1$ means finite, repulsive interactions.
%Here, we adopt the definition of the bosonic fields of Ref.~\onlinecite{Gia}, except for that the zero-modes $N_{+}$ and $N_{-}$ are written explicitly and are not included 
%in the fields $\phi$ and $\theta$ anymore.\\
From Eq.~\eqref{eq:Hll}, one can derive the bosonic form of the continuity equation 
\begin{equation}
\label{eq:dtphidxtheta}
\partial_t \phi - vK \partial_x \theta = 0\;.
\end{equation}
The fact that $vK = v_F(1 - g_2/(2\pi v_F))$ appears in this equation, instead of $v_F$, is a hallmark of 1D interacting Dirac fermions. Density operators actually do not commute, leading to a renormalization of the current operator by interactions. \\
 With this setup in mind, we now follow the lines of Ref.~\onlinecite{BaWiese03}, with the zero modes
unaffected by electron-electron interactions. We emphasize that this point will have important consequences on the transport properties, as we show in the following.
The idea behind this model is that the physical contacts attached to the quantum spin Hall edge are usually higher-dimensional, metallic objects of a greater extent than the edge channel itself. 
Since screening can work more efficiently here, we expect the leads to be barely interacting \cite{SafiSchulz, MaslovStone, Ponom95}. The full system is a composition of two external leads and the edge channel in between (see Fig.~\ref{fig:PlotLeads}). 
Thus, in the limit of large lead sizes, the zero modes of the full system will be dominated by the zero modes of the contacts \cite{BaWiese03}.  
Because of that, the influence of the non-interacting leads can be incorporated in the decoupled zero-modes.
%
% This can be seen as a pragmatical realization of a system with non-interacting semi-infinite leads and an interacting wire in between (see Fig.~\ref{fig:PlotLeads}). 
% In this model, we assume the total length of the system (with leads) to be much greater than the length of the wire only, $L$. 
In a second step, we then identify the length of the edge channel itself, $L$, as much greater than the remaining lengthscales of the system, implying the hierarchy $L\gg v_F/eV \gg a$.
Taking into account that position integrals can potentially compensate factors of $1/L$, we can apply the limit $L\to \infty$ to eventually drop the zero modes (and Klein factors).\\

To explore the underlying mechanisms, it is interesting to compare our findings to the well known problem of a regular impurity \cite{KaneFisher92}. Since at the quantum spin Hall edge backscattering is only possible with a simultaneous spin flip, 
the Rashba impurity $\Hr$ in this case is replaced by an impurity of the form  
\begin{align}\label{eq:DefHm}
 \Hm &=  \int dx\ m(x) \left( \PsdS{+}(x)\PsS{-}(x) + \Hc \right) \notag \\
 & = F_{+}^{\dagger} F_{-} \int dx \, \frac{m(x)}{2\pi a} \left(\frac{2\pi a}{L}\right)^K  \notag \\
 & \times \pp e^{2i (\phi(x)-\frac{\pi x}{L}(N_{+}+N_{-})) }  \pp + \Hc,
\end{align}
with impurity strength $m(x)$. Such an impurity can cause backscattering of a right-moving particle into a left-moving
particle, however, it does not preserve time-reversal symmetry, which is why we refer to it as a magnetic impurity.
Both perturbations will hinder electronic transport in a characteristic form.

\subsection{Non-equilibrium transport}

Transport signatures of the system are provided quite generally by quantities such as the average backscattering current or the current-current correlation (noise).  If a non-zero 
bias voltage is applied, this will modify the expectation value of any observable ${\mathcal{O}}$, for instance the current, to \cite{BaWiese03}
\begin{align}\label{eq:OaverageMain}
 \langle \mathcal{O} \rangle= \frac{1}{Z} \Tr(e^{-\beta \hVt} e^{i H t} \mathcal{O} e^{-i H t}).
\end{align} 
Here, we have defined $Z=\Tr(e^{-\beta \hVt})$, as well as $H=\Hll+\Hr$ and $\hVt=\Hll- \hV$.
As we describe in Appendix \ref{app:shifts}, the external bias can be implemented most conveniently by a unitary transformation $U$ on the free system, resulting in a shift of the bosonic fields. 
The expectation value of the observable $\mathcal{O}$ due to backscattering is then given by \cite{BaWiese03}
\begin{align}\label{eq:Oaverage4Main}
 & \langle {\mathcal{O}} \rangle= \frac{1}{Z_{LL}} \Tr(e^{-\beta \Hll} S^{\dagger}(t) \Oti(t) S(t)),
\end{align}
with
\begin{align*}
 S(t)=T \exp\left[-i \int_0^t dt' \left(\Hll+\Hti(t')\right)\right].
\end{align*}
The operator labels denote ${\tilde{\mathcal{O}}} =\Ud {\mathcal{O}} U$
and ${\mathcal{O}}_I(t)= e^{i \hV t} {\mathcal{O}} e^{-i \hV t}$ in the interaction picture. $U$ is the unitary operator specified in Eq.~(\ref{eq:DefU}) and $Z_{LL}=\Tr(e^{-\beta \Hll})$. 
The effect of external bias is thus fully absorbed in the two effective operators $\Hti(t)$ and $\Oti(t)$.\\ 
To calculate the expectation value in Eq.~(\ref{eq:Oaverage4Main}), we go beyond a linear-response model and use the Keldysh-Schwinger framework (for a review, see Ref.~\onlinecite{Rammer}). In such a closed time path formalism,
the statistical operator is designed to describe electronic transport in a system that is out of equilibrium due to an external perturbation \cite{Martin, Rammer}. 
The operator average takes then the form 
\begin{align}\label{eq:AvFormula}
& \langle {\mathcal{O}} (t) \rangle= \frac{1}{2} \sum_{\eta=\pm} \left\langle T_K \left[ \Oti^{\eta}(t) e^{-i \int_K dt_2 \Hti(t_2) } \right] \right\rangle_0.
\end{align}
The brackets $\langle \ldots \rangle_0$ denote the average with respect to the free Luttinger liquid Hamiltonian $\Hll$. $\eta=\pm$
is the Keldysh index representing the position of the respective operator on one of the two time branches of the Keldysh contour and $T_K$ denotes the Keldysh time-ordering operator \cite{Rammer}.
Here, we used $\int_K dt=\sum_{\eta'} \intinf dt\ \eta'$.
The expression can now be expanded up to the desired order in the external perturbation. In first order of the impurity Hamiltonian, it can be readily shown that the Keldysh time-ordering of the two operators
yields their retarded correlation function and Eq.~(\ref{eq:AvFormula}) becomes the usual Kubo formula \cite{Bruus}.
%
% \begin{align}\label{eq:Kubo0}
% & \langle {\mathcal{O}} (t) \rangle= 
% % \frac{1}{2} \sum_{\eta, \eta'=\pm} \left\langle T_K \left[ \Oti^{\eta}(t) (-\frac{i}{\hbar} \intinf \eta' dt_2 \Hti^{\eta'}(t_2)) \right] \right\rangle= \notag \\
% % &
% -i \intinf dt_2 \left\langle \left[\Oti(t), \Hti(t_2)\right]\right\rangle_0 \theta(t-t_2).
% \end{align}
% We used $\int_K dt=\sum_{\eta'} \intinf dt\ \eta'$.
% The Keldysh ordering for two operators in general corresponds to the retarded correlation function \cite{Bruus} 
% and Eq.~(\ref{eq:Kubo0}) is equivalent to the Kubo formula in first order of the impurity strength. 
%
In the following, let us define the transmitted current in the system as the difference of the free, and the backscattered current, $\jtr=\jfr-\jbs$. 
In this notation, $\jbs$ is a positive quantity, reducing the bare current due to backscattering off the impurity.
Finally, we can compute the correction to the conductance (resulting from the impurity) by
$ \delta G= \frac{d \langle \delta \jtr \rangle}{d V}= -\frac{d \langle \jbs \rangle}{d V}$, where we defined the correction to the transmitted current operator as $\delta \jtr=\jtr-\langle \jtr \rangle_0=-\jbs$.\\
As an additional quantity, the shot noise at zero frequency indicates the effective charge involved in the backscattering process. 
We calculate the symmetrized current-current correlation function \cite{Martin, Oreg}, which 
% \begin{align}\label{eq:Defnoise0}
%  S(t-t_2) & = 
% %  \sum_{\eta} \langle T_K \delta j^{\eta}(t)\  \delta j^{-\eta}(t_2) \rangle - 2 \langle \delta j \rangle^2= \notag \\ &
%  \sum_{\eta} \langle T_K  \jbs^{\eta}(t)\   \jbs^{-\eta}(t_2) \rangle - 2 \langle \jbs \rangle^2.
% \end{align}
at zero frequency reads
\begin{align}\label{eq:Defnoise}
 S(\omega \to 0)= \sum_{\eta} \int dt_2\ \langle T_K \delta {\jtr}^{\eta}(t)\  \delta {\jtr}^{-\eta}(t_2) \rangle .
%  -2 \langle \jbs \rangle^2 \delta(\omega).
\end{align}
The effective backscattered charge, also called Fano factor, is then given by the Schottky formula \cite{Oreg} $e^*=\frac{S}{2e |\langle \jbs \rangle|}$ in the weak backscattering limit.

\subsection{Effective operators}

The free current operator of the system simply reads
\begin{align}\label{eq:Defj}
& \jfr= \frac{e v_F}{L}(N_{+}-N_{-}).
% + \frac{evK}{\pi} \partial_x \theta(x,t),
% ev_F(N_{+}-N_{-})+\frac{e}{\pi}\partial_t \phi(x,t)= ev_F(N_{+}-N_{-})+\frac{i e}{\pi  }[\Hll,\phi(x,t)]= ev_F(N_{+}-N_{-})+ \frac{evK}{\pi  } \partial_x \theta(x,t),
\end{align}
% The continuity equation and the Heisenberg equation of motion yield $j(x,t)=\frac{1}{\pi}\partial_t \phi(x,t)= \frac{i}{\pi} [\Hll,\phi(x,t)]$, and we find for the bare current operator
% \begin{align}\label{eq:Defj}
% & j(x,t)= \frac{e v_F}{L}(N_{+}-N_{-})+ \frac{evK}{\pi} \partial_x \theta(x,t),
% ev_F(N_{+}-N_{-})+\frac{e}{\pi}\partial_t \phi(x,t)= ev_F(N_{+}-N_{-})+\frac{i e}{\pi  }[\Hll,\phi(x,t)]= ev_F(N_{+}-N_{-})+ \frac{evK}{\pi  } \partial_x \theta(x,t),
% \end{align}
% as is well known from the literature \cite{MaslovNotes}. 
Since the zero-modes are decoupled form the electron interactions, the bare Fermi velocity $v_F$ is implemented in Eq.~(\ref{eq:Defj}).
% , while the flow of bosonic excitations in the wire is not.
The external bias has the following impact (see Appendix \ref{app:shifts}), defining a new, effective free current operator
\begin{align*}
% & j(x,t)\to 
& \jfrti= \jfr + \frac{e^2 V}{2 \pi }.
\end{align*}
In the absence of impurities, we thus find from Eq.~(\ref{eq:Oaverage4Main}) the averaged transmitted current $\langle \jtr \rangle_0=\langle \jfrti \rangle_0=\frac{e^2 V}{2 \pi}$, representing the constant conductance $G_0=\frac{e^2}{2 \pi}$
of the helical edge channel. This contribution is interaction-independent, as it should be, which is a direct consequence of the decoupling of the zero-modes from the interactions in this model. 
% was proportional to $K$ if we would replace $v_F$ by $vK$ in Eq.~(\ref{eq:Defj})). 
It can be seen as a response of the system to the bias changing the number of right and left moving particles, independently of the system size. \\
To account for the influence of the impurity, we next define the backscattering current. It is given by 
% The observable of interest, however, to study transport in the presence of the Rashba impurity, is the backscattering current, 
the rate of changing the free current with time due to backscattering processes off the impurity\cite{Oreg}. 
Employing the Heisenberg equation of motion, 
% we have $\jbs\propto -\partial_t \jti \propto -\left[ \Hti,\jti \right]$. To make sure, that the backscattering current is again of dimension energy, 
we use \footnote{A factor of two is taken into account for the double counting of the charge, that is backscattered (compared e.g. to Ref.~12).}
\begin{align}
% \label{eq:Defjbs}
 \jbs(t)=-\frac{i}{2 v_F}\int dx \left[ \Hr (t),  \jfr \right]. \notag
\end{align}
%
% \section{Keldysh approach}
To compute the average backscattering current, we moreover need to find expressions for the operators $\jbsti$ and $\Hti$, modified by the voltage shift.
From now on, we specify the form of the impurity.
 For simplicity, we consider a point-like Dirac-impurity of the form $\alpha(x)=\alpha\ (a v_F)\ \delta(x)$, where
the impurity strength is given by the dimensionless parameter\cite{Crepin1} $\alpha$  (we take as well $m(x)= m\ {v_F} \delta(x) $ with dimensionless $m$ in Eq.~(\ref{eq:DefHm})). At this point, the limit $L\to \infty$ can safely be taken, 
and as a consequence, the zero modes $N_{\pm}$ are neglected. In the absence of zero modes, the Klein factors commute with 
all fields and will compensate in the averaged term, so we can drop them as well. We derive (see Eq.~(\ref{eq:shiftsOnHr})) 
\begin{widetext}
\begin{align}
\label{eq:HR3}
 \Hti (t) &=
i \frac{\alpha v_F}{\pi} \left(\frac{2\pi a}{L}\right)^K \left[
\pp \left(\partial_x \theta(0,t) + \frac{eV}{2 v_F} \right) e^{2i (\phi(0,t)+\frac{eV}{2} t) }  \pp +\Hc\right] \notag \\
&  =\frac{\alpha v_F}{2 \pi } \left(\frac{2\pi a}{L}\right)^K 
\left[ \frac{1}{vK} \pp \left(\partial_t e^{2i \phi(0,t)}\right) \pp e^{i eV t } + \frac{1}{v_F} \pp e^{2i \phi(0,t)} \pp \left(\partial_t e^{i eV t } \right) +\Hc\right].
% = \notag \\
%  &  \frac{\alpha v_F}{2 \pi} \left(\frac{2\pi a}{L}\right)^K 
% \left( \nu^{-1}(\partial_t)\ \partial_t \left( \pp e^{2i \phi(0,t)} \pp e^{i eV t }\right) +\Hc\right),
\end{align}
The structure of the backscattering current operator is very similar to the one of the impurity Hamiltonian, but note the opposite sign of the conjugate terms
% (The contribution was as well proportional to $K$ if we would replace $v_F$ by $vK$ in 
% Eq.~(\ref{eq:Defj})).
\begin{align}
\label{eq:jbsb}
 \jbsti(t) &= 
 -e \ \frac{\alpha v_F}{\pi} \left(\frac{2\pi a}{L}\right)^K
 \left[\pp \left(\partial_x \theta(0,t) + \frac{eV}{2 v_F} \right) e^{2i (\phi(0,t)+\frac{eV}{2} t) }  \pp -\Hc\right] \notag \\
 &  =i e \ \frac{\alpha  v_F}{2 \pi} \left(\frac{2\pi a}{L}\right)^K 
\left[ \frac{1}{vK} \pp \left(\partial_t e^{2i \phi(0,t)}\right) \pp e^{i eV t } + \frac{1}{v_F} \pp e^{2i \phi(0,t)} \pp \left(\partial_t e^{i eV t } \right) -\Hc\right].
%  & i e \ \frac{\alpha  v_F}{2 \pi} \left(\frac{2\pi a}{L}\right)^K 
% \left( \nu^{-1}(\partial_t)\ \partial_t \left( \pp e^{2i \phi(0,t)} \pp e^{i eV t }\right) -\Hc\right).
\end{align}
\end{widetext}
Since the impurity Hamiltonian is treated perturbatively here, we make use of the identity in Eq.~(\ref{eq:dtphidxtheta}). 
% At this point, we like to remind the reader, that we used the definition of fields following Ref.~\onlinecite{Gia}, that 
% sometimes differs by a sign compared to notations used by other authors \cite{Dolcini05,Budich}. 
Eqs.~(\ref{eq:HR3}) and (\ref{eq:jbsb}) will then be of the form of a total time
derivative in the cases of vanishing interactions $v\to v_F$ and $K\to 1$, Galilean invariance ($v K=v_F$), or a model with fully interacting leads. As we will see below, this corresponds each 
time to a zero backscattering current in lowest (second) order perturbation theory of the impurity strength. We observe from Eq.~(\ref{eq:jbsb}) that because of $\langle \delta \jtr \rangle_0 = -\langle \jbs \rangle_0=0$, there will be
% Eqs.~(\ref{eq:AvFormula}), ~(\ref{eq:HR3}) and ~(\ref{eq:jbsb})
no correction to the current in first order of the impurity strength. 
The same holds for all odd orders as a consequence of the ``neutrality rule'' for the respective correlation functions \cite{Gia}. In the following, we will evaluate the 
two lowest order contributions to the backscattering current, that are in second and fourth order of the impurity strength.

\section{Backscattering current to second order}
\subsection{Bosonic approach}
%
% We are interested in the average backscattering current depending on the external voltage, to study the effect of a Rashba impurity on the conductance of a helical Luttinger liquid.
%
Having derived the effective operators in the presence of the bias, we can now compute the backscattering current with the help of Eq.~(\ref{eq:AvFormula}).
To lowest (second) order of the impurity strength, we find (see Appendix \ref{app:2order})
\begin{align}\label{eq:avI1b}
 \langle \jbs \rangle &=
 \frac{1}{2} \sum_{\eta=\pm} \left\langle T_K \jbsti^{\eta}(t) (-i) \int_K dt_2 \Hti^{\eta'}(t_2) \right\rangle_0 \notag \\
% 
% p_1 \alpha^2 \left(\frac{2\pi a}{L}\right)^{2K} \int dt_2\ \theta(t-t_2) \nu^{-1}(\partial_t) \nu^{-1}(\partial_{t_2})\times \notag \\
%  & \bigg\langle \bigg[\left( \pp \partial_t\  e^{2i   \phi(0,t)+ i eV t}  \pp  - \Hc\right),
%     \left( \pp \partial_{t_2}\  e^{2i   \phi(0,t_2)+i eV t_2}  \pp  + \Hc\right) \bigg] \bigg\rangle_0 =\notag \\
% & 2i p_1 \alpha^2 \left(\frac{2\pi a}{L}\right)^{2K} \int d\tau\ \theta(\tau) \nu^{-2}(\partial_{\tau})\partial_{\tau}^2 \bigg( h(\tau) \sin(eV \tau) +h(-\tau) \sin(-eV \tau) \bigg) = \notag \\
% & 2i p_1 \alpha^2 \left(\frac{2\pi a}{L}\right)^{2K} \int d\tau  \nu^{-2}(\partial_{\tau}) \partial_{\tau}^2 \bigg( h(\tau) \sin( eV \tau) \bigg)= \notag \\
% & 2i p_1 \alpha^2 \left(\frac{2\pi a}{L}\right)^{2K} \frac{(Kv -v_F)^2}{(Kv\ v_F)^2} (-(eV)^2) \int d\tau   h(\tau) \sin( eV \tau)= \notag \\
% & \frac{-2 \pi p_1 \alpha^2}{\Gamma (2 K)} \left(\frac{a}{v}\right)^{2K} \frac{(Kv -v_F)^2}{(Kv\ v_F)^2}  (eV)^{2 K+1}= \\
 & =\frac{e\ \alpha^2 v_F^2}{2\pi \Gamma (2 K)} \left(\frac{a}{v}\right)^{2K} \frac{(Kv -v_F)^2}{(Kv\ v_F)^2}  (eV)^{2 K+1}.
\end{align}
The correction to the conductance in this order is thus $\delta G=-\frac{(2K+1) e^2\ \alpha^2 v_F^2}{2\pi \Gamma (2 K)} \left(\frac{a}{v}\right)^{2K} \frac{(Kv -v_F)^2}{(Kv\ v_F)^2}  (eV)^{2 K}$.
We note that in the non-interacting limit the average backscattering current vanishes, as it should be.
Importantly, we used the fact that the ground state is taken at zero temperature, corresponding to the regime $k_B T \ll eV$. The Poissonian shot noise can be evaluated analogously from Eq.~(\ref{eq:Defnoise}).
We obtain
\begin{align*}
 S(\omega\to 0)=  
% \sum_{\eta} \int dt_2\ \langle T_K \jbs^{\eta}(t)  \jbs^{-\eta}(t_2) \rangle = 
2e\ \langle \jbs \rangle.
\end{align*}
The effective backscattered charge in this case is $e^*=\frac{S}{2e \langle \jbs \rangle}=1$, indicating single-particle backscattering. \\
When comparing to the case of the magnetic impurity (see Eq.~(\ref{eq:DefHm})), we obtain in an identical fashion
\begin{align}
% \label{eq:avI1bMag}
& \langle \jbs \rangle_m=
% \frac{1}{2} \sum_{\eta=\pm} \left\langle T_K \jbsti^{\eta}(t) (-i) \int_K dt_2 H_m^{\eta'}(t_2) \right\rangle_0= \notag \\
  \frac{e\ m^2 v_F^2}{2\pi a^2 \Gamma (2 K)} \left(\frac{a}{v}\right)^{2K}  (eV)^{2 K-1}, \notag
\end{align}
and the same effective charge. 
In this case, we observe an additional factor of $V^{-2}$ compared to the Rashba impurity, due to the lack of the two spatial derivatives in the magnetic case.
Moreover, while the contribution of the Rashba SOC was primarily inelastic, the backscattering off the magnetic impurity does not vanish in the non-interacting limit.\\
The existence of a finite contribution in second order in Eq.~(\ref{eq:avI1b}) relies on the distinction of the two velocities $vK$ and $v_F$ in our setup. Such a distinction is made at two places, in
Eqs.~(\ref{eq:Hll}) and ~(\ref{eq:Defj}), associating $v_F$ with the zero-modes (modelling non-interacting leads) and $vK$ with the bosonic excitations.
At this point, we would like to investigate if such an assumption is in conflict with basic symmetries of the system, such as Galilean invariance. 
Indeed, as was originally pointed out by Haldane \cite{Haldane}, this symmetry requires $vK=v_F$, e.g. in systems with a parabolic energy dispersion.
We therefore expect that the second order contribution to the current in Eq.~(\ref{eq:avI1b}) would vanish if
Galilean invariance were to hold. 
The question of Galilean invariance is a well known problem of one-dimensional systems. 
As discussed in Ref.~\onlinecite{Starykh99}, 
this invariance is not automatically accounted for in Luttinger liquid theory, since it uses linearization of the energy spectrum as a vital requirement for bosonization. 
% However, Galilean invariance is manifest only in a parabolic dispersion \cite{Haldane}. 
When dealing with Luttinger liqids in an underlying parabolic system, one thus faces the problem of artificially broken Galilean invariance \cite{Haldane}. 
To cure this problem, different approaches were proposed, either defining new expressions for 
$v$ and $K$ to satisfy $vK=v_F$ \cite{Starykh99}, or by a careful treatment of higher harmonics of the density functions \cite{CappGia,Maslov05, ImamGlaz09a, ImamGlaz09b, ImamGlaz12}.
In the scenario described here, however, there is no need for such modifications, since Galilean invariance is emergently broken. 
Edge states in the gap of a topological insulator form a spectrum of two branches with opposite slope, that cross at some point in the bulk gap. Except for 
influences of remote bands, those branches are almost linear functions of the momentum, such that we find 
a solid ground for bosonization and the application of the Luttinger liquid model. The system is quasi relativistic, breaking Galilean invariance,
and therefore the Luttinger liquid relation $vK=v_F(1-g_2/(2\pi v_F))\neq v_F$ remains in general valid. Even though the second order correction to the conductance derived from Eq.~(\ref{eq:avI1b}) is still 
irrelevant in an RG sense, and thus never leads to localization in the low energy limit, we find in general a contribution scaling as $\delta G\sim V^2$ in the limit of weak interactions.
Studies of possible mechanisms to backscatter inelastically have revealed various power law corrections, indicating the inelasticity of the process \cite{Sch12, Budich, Oreg, Kain14, Stroem, Vay13}. Among such studies, the 
correction we describe above is one of the lowest powers ever found, and is thus expected to be of high relevance for transport.\\
%
%
% To support our result, we performed an additional calculation in fermionic language using the generalized Fermi's golden rule (see supplementary material). The analog of non-interacting leads
% would here be the exclusion of the elastic terms from the interaction Hamiltonian. With this assumption we find a backscattering current contribution in second order of both the impurity and the interaction 
% strength $g_2$, that scales approximately as $V^3$, as expected from Eq.~(\ref{eq:avI1b}) in the limit of weak interactions. Physically, such processes correspond to an inelastic backscattering of a single particle accompanied by 
% a particle-hole excitation at one of the branches.\\
%
If Galilean invariance was restored for any reason, or interactions were fine-tuned such that $vK=v_F$, the SPB contribution would vanish, and the lowest order correction to the conductance would be given by correlated two-particle 
processes, as discussed in the Sec.~\ref{sec:4order} below.

\subsection{Fermionic approach}
\label{sec:Fermions}
To support our result, we performed an additional calculation in fermionic language using the generalized Fermi's golden rule. According to this model, 
the average inelastic backscattering current is proportional to the transition rate \cite{Kain14, Sch12} $| \langle f | T | i \rangle|^2 \delta_{\epsilon_i, \epsilon_f}$, where 
$|i\rangle$ and $|f\rangle$ describe intial and final multi-particle states that, in the case of single-particle backscattering, differ by one in the number of right and left moving particles. 
$\epsilon_i$ and $\epsilon_f$ are the energies of the two respective states, and the 
T-matrix operator is given by \cite{Bruus} $T= H' + H' G_0 T$. We specify the perturbation Hamiltonian $H'=H_R+\Hint$ and $G_0=\frac{1}{\epsilon_i-H_0}$, where $H_0$ is the free 
kinetic Hamiltonian.
The average backscattering current is precisely defined as
%\cite{Budich, Crepin1}
% \begin{align*}
%  I_{BS}=\frac{e L^2}{4\pi^2} \sum_{k_i,k_f}n(\epsilon_i) (1-n(\epsilon_f)) |M_{if}|^2 \delta(\epsilon_i-\epsilon_f)
% \end{align*}
% \begin{align}\label{eq:Ibs}
%   I_{BS}&=e \frac{L^4}{(2\pi)^4} \int dk_i dk_j dk_f dk_g \sum_{\eta_i,\eta_j,\eta_f,\eta_g} n(\epsilon_i)n(\epsilon_j) \bar{n}(\epsilon_f) \bar{n}(\epsilon_g) |M_{ij;fg}|^2 \delta(\epsilon_i+\epsilon_j-\epsilon_f-\epsilon_g)=\notag \\
%  & \frac{e L}{v_F} \sum_{i,j,f,g} n(\epsilon_i)n(\epsilon_j) \bar{n}(\epsilon_f) \bar{n}(\epsilon_g) |M_{ij;fg}|^2 \delta_{k_i+k_j,k_f+k_g}
% \end{align}
%
\begin{align}\label{eq:Ibs}
 \langle \jbs \rangle &= \frac{e L}{v_F} \sum_{f} |M_{if}|^2 \delta_{\epsilon_i/v_F,\epsilon_f/v_F}.
\end{align}
% Here, $|i\rangle$ and $|f\rangle$ describe an intial and final multi-particle state to be specified later. 
Here, $M_{if}$ is the transition matrix element, given by $M_{if}= \langle f | T | i \rangle$.
%  The T-matrix operator is now given by \cite{Bruus} $T= H' + H' G_0 T$, where we specify $H'=H_R+\Hint$ and $G_0=\frac{1}{\epsilon_i-H_0}$.
%  Hereby, the free Hamiltonian reads $H_0=\frac{1}{L} \sum_k \xi_{+}(k) \Psd{+}{k}\Ps{+}{k}+\xi_{-}(k) \Psd{-}{k}\Ps{-}{k}$, with the single-particle energies $\xi_{\pm}(k)$.\\
%
 We focus on single-particle contributions, where $M_{if}$ is of order $\alpha\ g_2$. Therefore, we go
 to second order in $T=H' G_0 H'$ and consider the terms 
 \begin{align}\label{eq:Mif}
 M_{if}=\langle f| H' G_0 H' | i\rangle = \langle f | H_R G_0 \Hint+ \Hint G_0 H_R | i \rangle. 
%  \equiv (I)+(II).
\end{align}
We next derive the form of the effective operator $T$ in momentum space, where we define 
for the Fourier components\footnote{From this definition we infer that $\Psd{\pm}{k}$ is dimensionless, since $\PsdS{\pm}(x)$ has dimension $L^{-1/2}$.}
$\Psi^{\dagger}_{\pm}(x)=\frac{1}{\sqrt{L}}\sum_k e^{+ ikx}\Psd{\pm}{k}$.
The free Hamiltonian can thus be written as $H_0= \sum_k \xi_{+}(k) \Psd{+}{k}\Ps{+}{k}+\xi_{-}(k) \Psd{-}{k}\Ps{-}{k}$ with single-particle energies $\xi_{\pm}(k)=\pm v_F k$
for the right and left movers.\\
The impurity Hamiltonian in momentum space becomes
\begin{align}
 \Hr &=\int dx\ \alpha(x)\left[ \left(\partial_x \Psi^{\dagger}_+\right) \Psi_- - \Psi^{\dagger}_+ \left(\partial_x \Psi^{\phantom{\dagger}}_- \right) \right](x) + \Hc \notag \\
%  &  \sum_{k_a,k_b} \Gamma_{ab} \Psd{+}{k_a}\Ps{-}{k_b} + \Gamma_{ba} \Psd{-}{k_b}\Ps{+}{k_a}= \notag \\
& =\sum_{k_a,k_b} \Gamma_{ab} (\Psd{+}{k_a}\Ps{-}{k_b} - \Psd{-}{k_b}\Ps{+}{k_a}).
\end{align}
with $\alpha(x)=\alpha L v_F\ \delta(x)$ and $\Gamma_{ab}={i\alpha v_F}{}(k_a+k_b)$.
% , $\Gamma_{ba}={-i\alpha v_F}{}(k_a+k_b)=-\Gamma_{ab}$.
The interaction Hamiltonian on the other hand reads
\begin{align}
 \Hint &= -g_2 \int dx\ \Psi^{\dagger}_{+}(x) \Psi^{\dagger}_{-}(x) \Psi_{+} (x)\Psi_{-}(x) \notag \\
 & =\frac{-g_2}{L} \sum_{k_r,k_s,q}\Psd{+}{k_r-q} \Psd{-}{k_s+q} \Ps{+}{k_r}\Ps{-}{k_s}.
\end{align}
In this notation, $\alpha$ is a dimensionless parameter, while $g_2$ has the dimension of a velocity.
Now, we can compute the two parts of $T$ in Eq.~(\ref{eq:Mif}). Using the anticommutation relation $\{\Ps{\eta_i}{k_i},\Psd{\eta_j}{k_j} \}=\{\PsS{i},\PsdS{j} \}=\delta_{i,j}= \delta_{k_i, k_j}\delta_{\eta_i, \eta_j}$,
we find after some algebra
 \begin{align}\label{eq:Teffective}
  T|i\rangle &= 
%   \beta \sum_{k_a,k_b, k_s, q} \left(\bigg[\Psd{-}{k_a-q}\Psd{+}{k_s+q}\Ps{+}{k_s}\Ps{+}{k_b} - \Psd{-}{k_a}\Psd{-}{k_s-q}\Ps{-}{k_s}\Ps{+}{k_b-q}\bigg] - \{ + \leftrightarrow - \}\right)|i\rangle=\notag \\
    \beta \sum_{k_a, k_s, q} \bigg(\bigg[\Psd{-}{-k_a+q}\Psd{+}{k_s+q}\Ps{+}{k_s}\Ps{+}{k_a}  \notag \\
  & - \Psd{-}{-k_a+q}\Psd{-}{-k_s-q}\Ps{-}{-k_s}\Ps{+}{k_a}\bigg] - \{ + \leftrightarrow - \}\bigg)|i\rangle,
  \end{align}
with $\beta=\frac{-i\alpha g_2}{L}$. 
 Here, we made use of energy conservation, requiring $-(k_a-q)+(k_s+q)-k_s-k_b=0 $, meaning that by setting $k_b=2q-k_a$, the sum over one momentum can be dropped
 \footnote{Subsequently, we performed the relabeling $k_a \to k_a +2q$ in the first term and $k_a \to k_a +q$ in the second, and changed for both the notation $k_a \to -k_a$, as well as $k_s \to -k_s$ in the second term.}.
 As expected, the energy denominator cancelled with the momenta in $\Gamma_{ab}$, regularizing the expression. \\
% For the second term in Eq.~(\ref{eq:Mif}), we can apply $G_0$ to the left, since $\epsilon_i=\epsilon_f$ for both states, though we gain an additional sign because of the complex conjugate of the Rashba.\\
%
%
Let us briefly interpret Eq.~(\ref{eq:Teffective}). 
Thinking of $\sum_{k_s}\Psd{\pm}{k_s \pm q}\Ps{\pm}{k_s}=\rho_{\pm,\pm q}^{\dagger}$ as a particle density, the effective operator $T$ exhibits the same structure as the bosonized Rashba Hamiltonian in the 
first line of Eq.~(\ref{eq:HR3}). It describes a backscattering process 
 coupled to a bosonic excitation at one of the branches and, importantly, implies finite interactions. 
  Without interactions, we can readily check that the transition matrix elements $\langle f | H_R | i\rangle$ and $\langle f | H_R G_0 H_R | i\rangle$ are always zero because
of energy conservation. This result embodies the non-interacting limit and was shown already in Ref.~\onlinecite{Crepin1}.\\
 The components with $q=0$ in Eq.~(\ref{eq:Teffective}) can be identified with the term $(N_+-N_-)e^{2i\phi(0)}\sim Ve^{2i\phi(0)}$ in Eq.~(\ref{eq:HR3}), and thus 
 manifest the zero-mode part of the T-matrix. 
 In the bosonic calculation, we assumed that the zero-modes, representing the external contacts, were decoupled from the 
 electron interactions. To adopt the same model here, we have to impose an additional constraint on the effective 
 operator in Eq.~(\ref{eq:Teffective}) in such a way, that the interacting, effective operator does not alter the zero-modes. 
 This can be achieved by excluding the $q=0$ components explicitly from Eq.~(\ref{eq:Teffective}). 
 Indeed we find from an 
 explicit calculation (see Appendix \ref{app:Fermionic}), that $M_{if}=0$ if all values of $q$ are taken into account in Eq.~(\ref{eq:Teffective}). This case corresponds to the situation when 
 interactions do affect the contacts, and the finding is thus in agreement with Eq.~(\ref{eq:avI1b}). 
%  The cancellation mechanism is, that one term with nonzero $q$ will always be compensated by another one with $q$ equals zero in any of the terms given in Eq.~(\ref{eq:Teffective}).
 On the other hand, if we exclude the components with $q=0$, we get $M_{if}\neq 0$ and a finite contribution in second order of the impurity strength. This case reflects the 
 setup we consider in this article, with the leads decoupled from the electron interactions.\\
%  Both findings are in agreement with the result in Eq.~(\ref{eq:avI1b}).
%
%
Defining a system with a finite, effective bandwidth, we numerically evaluate Eq.~(\ref{eq:Ibs}). Hereby, we fix $|i\rangle$ as an initial state of the system in the presence 
of an external bias, and sum over all final states $|f\rangle$. 
In both the initial and the final state, we account for a Fermi sea of finite depth, and finally estimate the dependency of $\jbs$ on the voltage bias (see Appendix \ref{app:Fermionic}).
We conclude, that the backscattering current can be approximately given by the expression 
\begin{align}\label{eq:IbsFinalMain}
 \langle \jbs \rangle \sim e \alpha^2 g_2^2 \frac{L^2}{v_F^4(2\pi)^3} (eV)^3,
\end{align}
which is in qualitative agreement with the result in Eq.~(\ref{eq:avI1b}), in the limit of weak interactions.

\section{Backscattering current to fourth order}
\label{sec:4order}

The analytical calculation of the average backscattering current in fourth order is a challenging task, since in the Keldysh scheme, multi-dimensional coupled integrals are encountered. 
In principle, it will turn out to be very tedious if not even impossible to account for all the contributions arising. Such contributions can quite generally be 
processes involving two particles that are coupled by interactions. For simplicity, we consider here only the two extreme limits of the problem. 
Those are the cases of strongly coupled two-particle backscattering and the backscattering of two decoupled particles. 
Technically, these limits are taken by performing contractions on the time variables. The remaining integrals can then be solved to reveal the respective contributions to 
the backscattering current (see Appendix \ref{app:4order}). \\
We first focus on the strongly coupled TPB limit. It is of particular importance, since TPB at strong interactions was shown to be a potentially relevant perturbation in the RG sense, leading 
to localization and a breakdown of the conductance in the low energy limit\cite{Stroem,Crepin1,GeissCrep14}.
%
% Since the main focus in this order is on the two-particle backscattering terms, 
% we therefore apply contractions on the time variables, to source out the respective terms and solve the remaining integrals (see supplementary material). 
%
The TPB contractions are only meaningful if the processes selected in the procedure
represent the most important contributions of all processes. In particular, this requires that the time variables we consider as small in the contraction process appear indeed with sufficiently fast decaying power laws.
For the Rashba impurity, this is the case only in the regime $K<1/2$.
For such strong interactions, we can infer that the TPB processes are of leading power in the bias and we find (Appendix \ref{app:4order})
\begin{align}\label{eq:jbsRKsm}
 \langle  \jbs \rangle \sim
  \frac{\pi  2^{8 K+5} }{K^2 \Gamma (8 K)} \frac{e \alpha^4}{4 (2 \pi)^4} \left(\frac{a}{v}\right)^{8K-2} \left(\frac{v_F}{v}\right)^4 (eV)^{8 K-1},
\end{align}
if $K<1/2$.
% and $\delta G= \frac{(8K-1)\pi  2^{8 K+5} }{K^2 \Gamma (8 K)} \frac{e \alpha^4}{4 (2 \pi)^4} \left(\frac{a}{v}\right)^{8K-2} \left(\frac{v_F}{v}\right)^4 (eV)^{8 K-2}$.
Likewise, the noise can then be derived as
\begin{align}
 S(\omega\to 0) = 4e \langle  \jbs \rangle. \notag
\end{align}
We hence obtain an effective charge $ e^{*} = 2$. This result, in addition to the expected scaling of $V^{8K-1}$, indeed confirms a pure TPB process \cite{Stroem, Crepin1, Oreg, MaciejkoOreg09}.\\
Studying the same contraction procedure for the magnetic impurity, we see that these TPB contractions are never justified, because of the lack of derivatives in this case. Therefore, terms as found 
in Eq.~(\ref{eq:jbsRKsm}) are never the leading contributions. 
The TPB current given above survives in the case of restored Galilean invariance, $vK=v_F$, in contrast to the SPB processes of Eq.~(\ref{eq:avI1b}). 
In an experiment at $K<1/2$, we should find $ e^{*} = 2$ from TPB in the presence of Galilean invariance, while in the absence of this symmetry, the second order SPB contribution was dominant, leading to $e^*=1$. 
Consequently, besides the distinct power scaling, we can identify the Fano factor as a direct evidence for the presence of Galilean invariance, which should therefore be experimentally measurable. \\
% With Galilean invariance and strong interactions $K<1/2$, we expect a Fano factor of two, whereas in the absence of this symmetry, the lowest order contribution yields a Fano factor of one.\\
%
%
Next, we consider the situation of weak interaction strengths $K\geq 1/2$. In this regime, the TPB contractions used before are not well justified, since they do not capture the essential subset of processes any more.
% First, at the point $K=1/2$, the average TPB current is found to be vanishing.
However, we can still observe the change of the scaling of the strongly coupled TPB terms. When going from $K<1/2$ to $K>1/2$, we infer from the 
structure of the encountered integrals that the scaling of such terms is changed from $V^{8K-1}$ to $V^{4K+1}$. We have to keep in mind though, 
that the leading contributions for $K>1/2$ are not represented by pure TPB processes any more, but should be more of the form of two weakly coupled backscattering events.
% , keeping in mind though, that those do not represent the leading contribution any more for $K>1/2$.
%
% Following still the same contraction procedure for $K>1/2$, we can check how the strongly coupled TPB terms scale in this regime.
% % At the point $K=1/2$, the average TPB current is found to be vanishing, while for $K>1/2$, the TPB contractions are not well justified, since they do not capture the essential subset of processes any more. 
% We find that the scaling of such terms is now changed to $V^{4K+1}$, keeping in mind though, that those do not represent the leading contribution any more.
% The truly leading terms should be more of the form of two weakly coupled backscattering events, but we have no contraction procedure at hand to exactly source them out. 
% However, from the structure of the encountered integrals we infer that the scaling of those terms can as well only be $V^{4K+1}$, meaning that 
% the TPB contraction, which is not well justified for $K>1/2$, should still give a correct idea of the power scaling.
%
Along this line, we can confirm the existence of a crossover of scales at $K=1/2$, for correlated two-particle processes, as 
discovered in Ref.~\onlinecite{Crepin1}.\\
Finally, to complement our fourth-order analysis, we consider the opposite limit of two processes decoupling into separate single-particle events.
Here, we perform the contraction of time variables in a different way to sort out the terms of interest (see Appendix \ref{app:4order}). We then find 
% To better understand the character of the backscattering processes, it seems interesting to contract in a different way, to reveal single-particle contributions (see Appendix~\ref{app:4order}). 
\begin{align}\label{eq:jbsRKla}
 & \langle \jbs \rangle  \sim  \frac{(v K-v_F)^4}{(vK)^4} \frac{\alpha^4 e}{16 \pi^2 \Gamma (2 K)^2} \left(\frac{a}{v}\right)^{4K+1} (eV)^{4 K+2},   \\
 & S(\omega\to0)  \sim 4 e \langle \jbs \rangle. \notag
\end{align}
% The correction to the conductance in this case becomes $ \delta G=\frac{(4K+2)(v K-v_F)^4}{(vK)^4} \frac{\alpha^4 e}{16 \pi^2 \Gamma (2 K)^2} \left(\frac{a}{v}\right)^{4K+1} (eV)^{4 K+1}$
In this case, the contribution has a similar structure as the backscattering off a magnetic impurity, where we can derive
\begin{align}\label{eq:SPBcontr}
 & \langle  \jbs \rangle_m \sim \frac{m^4 e}{16 \pi^2 \Gamma (2 K)^2} \left(\frac{a}{v}\right)^{4K-3} \left(\frac{v_F}{v}\right)^4 (eV)^{4 K-2},  
\end{align}
and the same Fano factor. Since two times a single charge is transfered by the two decoupled SPB events, we find an effective charge of two, meaning 
that we can not distinguish double SPB and TPB from the noise only, but from the scaling with the bias. 
This kind of SPB contraction is always justifiable, which is why the current contribution in Eq.~(\ref{eq:jbsRKla}) holds for all $0<K<1$, however, 
the TPB terms will still dominate for strong Coulomb interactions. 
% since we take into account small times of
% the functions decaying with the power $−2K$. This fact translates into a well-behaved expression with no singularities. The expression
% does not vanish for $K \to 1$, meaning that in this limit and fourth order of the impurity strength, SPB is present for the magnetic impurity, but not in the Rashba impurity case because of 
% the prefactor in Eq.~(\ref{eq:jbsRKla}).
Consistently, the SPB scaling of $V^{4K+2}$ exhibits exactly one more power of $V$ than the above mentioned correlated two-particle processes for $K>1/2$. The missing power in the latter 
case reflects the existence of one additional time integral, creating a weak link between the two events. \\
With Eqs.~(\ref{eq:jbsRKsm}) and (\ref{eq:jbsRKla}), we have given the two interesting limits in this order of $\alpha$, being strongly coupled two-particle backscattering and decoupled double single-particle 
backscattering, respectively. The intermediate regime can not be captured exhaustively in our present analysis, though a transition between the two limits is indicated 
by the power scalings found from the structure of the integrals.

\section{Conclusion}

In this article, we have studied the influence of a local Rashba spin-orbit scatterer on the edge transport of a helical quantum spin Hall system. 
In contrast to previous analyses, we have applied a non-equilibrium approach with explicit implementation of the contacts attached to the sample. Indeed we have found that the modeling of the leads
is of qualitative importance for the resulting backscattering current in lowest order. Because an emergent Lorentz invariance is characteristic of the 
QSH edge channels, we have discovered a single-particle backscattering allowed at second order of the impurity strength. This contribution differs from the backscattering off a TRS-breaking impurity by its 
purely inelastic character and by the power law scaling.\\
Subsequently, we have extended our calculation up to fourth order in the impurity strength, to verify the existence of correlated TPB processes. As expected we have identified the corresponding 
power law contributions of the correction to the conductance, as well as the crossover of scalings at $K=1/2$, in agreement with Ref.~\onlinecite{Crepin1}. The analysis of the shot noise has yielded the 
backscattered charges in the limits of pure SPB or TPB. The resulting Fano factor of one or two, respectively, can be seen as a direct evidence for the absence or presence of Galilean invariance.\\
We thank R. Egger, N. Kainaris and T. Schmidt for helpful discussions.
Financial support by the DFG (German-Japanese research unit ``Topotronics'', the priority program SPP 1666, and the SFB 1170 ``ToCoTronics'', as well as the Helmholtz Foundation (VITI), and the ``Elitenetzwerk Bayern`` 
(ENB graduate school on ``Topological insulators'') is gratefully acknowledged.

\newpage
%---------------------------------------------------------------------------------------------------------------------------------------%
% \section{The HLL-SC interface}

\newpage
\appendix
\onecolumngrid

\section{Form of the interaction Hamiltonian}
\label{app:interactions}

Let us comment here on the specific form of electron-electron interactions in a helical liquid.
In a usual, spinful Luttinger liquid, Coulomb interactions are generally described by two-body interactions of the form 
$\Hint= \int dx \int dy\ \Psid(x) \Psid(y) U(x-y)\Psi (y) \Psi (x)$ \cite{Haldane, Haldane2, Bruus, Schoen97}, with operators arranged in normal order. 
$\Psid(x)$ and $\Psi(x)$ are fermionic creation and annihilation operators for an electron at position $x$, that will be specified with additional indices for momentum and spin. 
In general, at the QSHE edge, we expect only short range ineractions due to screening. 
Assuming contact interactions, $U(x-y)=U_0\ \delta(x-y)$, we can classify three types of interactions $H_{int}=H_1+H_2+H_4$, with
% \begin{align*}
%  & \Hint= H_1+H_2+H_4, \\
%  &H_4= \sum_{\sigma,\sigma'=\uparrow,\downarrow} \int dx \int dy U(x-y) \psid_{+,\sigma}(x) \psid_{+,\sigma'}(y) \psi_{+,\sigma'} (y) \psi_{+,\sigma} (x) + \{+ \leftrightarrow -\},\\
%  & H_2= \sum_{\sigma,\sigma'=\uparrow,\downarrow} \int dx \int dy U(x-y) \psid_{+,\sigma}(x) \psid_{-,\sigma'}(y) \psi_{-,\sigma'} (y) \psi_{+,\sigma} (x)+ \{+ \leftrightarrow -\},\\
%  & H_1=\sum_{\sigma,\sigma'=\uparrow,\downarrow} \int dx \int dy U(x-y) \psid_{+,\sigma}(x) \psid_{-,\sigma'}(y) \psi_{+,\sigma'} (y) \psi_{-,\sigma} (x)+ \{+ \leftrightarrow -\}.
% \end{align*}
%+ \{+ \leftrightarrow -\}
\begin{align}\label{eq:H1}
 & H_1=U_0 \sum_{\sigma,\sigma'=\pm 1/2} \int dx\  \Psid_{+,\sigma}(x) \Psid_{-,\sigma'}(x) \Psi_{+,\sigma'} (x) \Psi_{-,\sigma} (x), \\
 \label{eq:H2}
  & H_2= U_0\sum_{\sigma,\sigma'=\pm 1/2} \int dx\  \Psid_{+,\sigma}(x) \Psid_{-,\sigma'}(x) \Psi_{-,\sigma'} (x) \Psi_{+,\sigma} (x),\\
  \label{eq:H4}
   &H_4= U_0 \sum_{\sigma,\sigma'=\pm 1/2} \int dx\  \Psid_{+,\sigma}(x) \Psid_{+,\sigma'}(x) \Psi_{+,\sigma'} (x) \Psi_{+,\sigma} (x).
\end{align}
Here, we have specified the fermionic field operators such that 
$\Psi^{\dagger}_{\pm,\sigma}(x)$ creates a right $(+)$ or left $(-)$ moving particle with spin $\sigma=\pm 1/2$. In the helical liquid, because 
of spin-momentum locking, one index is redundant and can be dropped again later. 
% However, we keep the index for a moment out of pedagogical reasons.
In terms of the usual g-ology \cite{Starykh99, Maslov05, Gia}, we could replace, in each of the Eqs.~(\ref{eq:H1}) to (\ref{eq:H4}),
$U_0$ by prefactors $g_{1,\parallel/\bot}=g_{2,\parallel/\bot}=g_{4,\parallel/\bot}=U_0$, where the index $\parallel$ indicates parallel spins $\sigma=\sigma'$ and 
$\bot$ represents $\sigma=-\sigma'$.\\
Here, the Pauli principle is manifested in forbidding terms $g_{4,\parallel}$, since in Eq.~(\ref{eq:H1}) $\psi_{+,\sigma} (x) \psi_{+,\sigma}(x)=0$. Moreover, we see that 
after fermionic anticommutation $g_{1,\parallel}$ is of the same form as $g_{2,\parallel}$ but with a minus sign, and therefore cancel each other in case of the contact interactions we assumed here. 
Because of this, all the parallel interactions are absent.\\
In the helical liquid, we have a special situation because of spin-momentum-locking. This tells us, that processes $g_{4,\bot}, g_{1,\bot}$ (and $g_{1,\parallel}, g_{2,\parallel} $)
are impossible due to the helical character of the edge states, therefore, the only remaining interaction term is $g_{2,\bot}$, which is
\begin{align*}
 & H_2= U_0\sum_{\sigma=\uparrow,\downarrow} \int dx  \Psid_{+,\sigma}(x) \Psid_{-,-\sigma}(x) \Psi_{-,-\sigma} (x) \Psi_{+,\sigma} (x).
\end{align*}
Since spin and momentum are locked, one of the two indices can safely be dropped. Importantly, 
the helical liquid is only formally spinless, but as we pointed out, there is an essential difference between the helical and the spinless case, when it comes to 
Coulomb interactions. In a spinless picture, we would expect no contribution at 
all in case of contact interactions, since $g_4$ cancels due to the Pauli principle and as well $g_2-g_1=0$ \cite{KaneFisher92, Maslov05}. From these considerations, for contact interaction, we have in the helical liquid
\begin{align*}
& \Hint= g_2 \int dx\ \Psi^{\dagger}_+(x) \Psi^{\dagger}_-(x) \Psi_-(x)\Psi_+(x).
\end{align*}

\section{Derivation of the shift of bosonic fields}
\label{app:shifts}

Starting from Eq.~(\ref{eq:OaverageMain}), we rederive here the shift of fields as presented in Eq.~(\ref{eq:Oaverage4Main}), following schematically Ref.~\onlinecite{BaWiese03}.
To do so, we introduce an unitary operator $U$, defined as
\begin{align}\label{eq:DefU}
 U=e^{i \alpha V (f_+ - f_-)}.
\end{align}
$\alpha$ is a constant factor not yet specified. Here, the Klein factors are written in exponential form, $F^{\dagger}_{\pm}=e^{i f_{\pm}}$ and $F^{}_{\pm}=e^{-i f_{\pm}}$. Those 
Klein factors commute with the bosonic creation and annihilation operators, though not with the particle number operators. 
The commutation relations read $[N_{\eta},f_{\eta'}]= -i\delta_{\eta \eta'}$, and $[f_{\eta}, f_{\eta'}]=[N_{\eta}, N_{\eta'}]=0$ with $\eta=\pm$ \cite{BaWiese03,Delft}.
We now see that $U$ is able to establish the shift
\begin{align}
% \label{eq:UHllU}
 & U e^{-\beta \Hll} \Ud= e^{-\beta (\Hll - \hV)} e^C= e^{-\beta \hVt} e^C, \notag
\end{align}
% \begin{align*}
% & U e^{-\beta \Hll} \Ud= 
% % e^{-\beta \Hll} U e^{-\beta [i \alpha V \int dx \phi(x), \frac{v}{2\pi}\int dx' K (\partial_{x'} \theta(x'))^2]} \Ud= \\
% %  &  e^{-\beta \Hll} U e^{-\beta (-\alpha V K v) \int dx \partial_{x} \theta(x)} \Ud= e^{-\beta \Hll} e^{-\beta (-\alpha V K v) \int dx \partial_{x} \theta(x)} e^C U \Ud=\\
% %  & e^{-\beta \Hll} e^{-\beta (-\alpha V K v) \int dx \partial_{x} \theta(x)} e^C = \\
% %  & e^{-\beta (\Hll+ (-\alpha V K v) \int dx \partial_{x} \theta(x))} e^{\frac{1}{2}\beta^2 [\Hll,-\alpha V K v \int dx \partial_{x} \theta(x) ] } e^C = \\
%   e^{-\beta (\Hll-\frac{4\alpha \pi v_F}{L}  \hV) } e^C .
% %  & e^{-\beta (\Hll -\frac{\alpha v}{\omega}  \hat{V}) } e^{-i \frac{\beta^2 v^2 \alpha V}{2}\int dx \partial_x^2 \phi} e^C
% \end{align*}
when $\alpha= eL/(4v_F\pi)$ is chosen. $e^C= e^{4\beta V^2 \alpha^2 \pi v_F/L}$ is just a constant created by commutating $U$ past $e^{\beta \hV}$. 
% It is useful to note, that (trivially) the commutator $ [\Hll,\hV] =0$ drops out.
%
Applying $\Ud$ from the left hand side is equivalent to a voltage-dependent shift on the particle number operators $N_{\pm} \to N_{\pm} \pm \frac{eV L}{4 \pi v_F}$.
With this, we write now Eq.~(\ref{eq:OaverageMain}) as
\begin{align}\label{eq:Oaverage2}
 & \langle {\mathcal{O}} \rangle= \frac{1}{Z_{LL}} \Tr(U e^{-\beta \Hll} \Ud e^{i H t} {\mathcal{O}} e^{-i H t}) \notag \\
%  & \frac{1}{Z_{LL}} \Tr(e^{-\beta \Hll} (\Ud e^{i {H} t} U)(\Ud {\mathcal{O}} U)(\Ud e^{-i {H} t} U)).
 & \sim \frac{1}{Z_{LL}} \Tr(e^{-\beta \Hll} e^{i (\Hll +\hV+ \Ht) t} \tilde{{\mathcal{O}}} e^{-i (\Hll +\hV+ \Ht) t}).
\end{align}
with $Z_{LL}=\Tr(e^{-\beta \Hll})$ and ${H}= \Hll+\Hr$. Moreover, we define 
for a general operator $\tilde{{A}}\equiv (\Ud {A} U)$. Above, we used the cyclic permutation of the trace and inserted factors of $U\Ud=1$. 
Since the system is mainly governed by the exponents linear in time, we approximately neglected all the terms of the power of $t^2$, leading to $\Ud e^{iHt}U \sim e^{i (\Hll +\hV+ \Ht) t} e^{C/\beta^2}$.
Next, we go to an interaction picture \cite{BaWiese03}, using that for any operator $A_I(t)= e^{i \hV t} A\ e^{-i \hV t}$, we write
\begin{align*}
 e^{-it (\Hll + \Ht+\hV )}= e^{-it \hV } T \exp\left[-i \int_0^t dt' \Hll(t')+\Hti(t')\right] \equiv e^{-it \hV} S(t).
\end{align*}
Eq.~(\ref{eq:Oaverage2}) then becomes
\begin{align}\label{eq:Oaverage4}
 & \langle {\mathcal{O}} \rangle=\frac{1}{Z_{LL}} \Tr(e^{-\beta \Hll} S^{\dagger}(t) e^{it \hV} \tilde{{\mathcal{O}}} e^{-it \hV} S(t))= 
 \frac{1}{Z_{LL}} \Tr(e^{-\beta \Hll} S^{\dagger}(t) \Oti(t) S(t)),
\end{align}
and we arrive at Eq.~(\ref{eq:Oaverage4Main}) of the main part.
% (Cf. result to Ref.~\onlinecite{BaWiese03}, Eq.~(13)).
In brief, the result in Eq.~(\ref{eq:Oaverage4}) means, that introducing a finite voltage of the above form is equivalent 
to transforming both the perturbative impurity Hamiltonian $\Hr \to \Hti(t)$ and the observable under consideration $\mathcal{O} \to \Oti(t)$.\\
% , where we defined in general
% \begin{align}\label{eq:ObservableShift}
% & \Ati (t)= e^{+i \hV t} (\Ud {A}\ U) e^{-i \hV t}.
% \end{align}
The shift of the bosonic fields now depends on the form of the operators appearing in the system. In our case we find $\phi(x)\to \phi(x)+ \frac{eV}{2} t$ and 
$\partial_x \theta(x) \to \partial_x \theta(x) + \frac{eV}{2 v_F}$.
% \begin{align}
% & \phi(x)\to \phi(x)+ \frac{eV}{2} t, \notag \\
% & \partial_x \theta(x) \to \partial_x \theta(x) + \frac{eV}{2 v_F}. \notag
% \end{align}
(cf. Ref.~\onlinecite{Oreg}). 
% A rescaling of fields $\phi\to \sk \phi$ and $\partial_x \theta \to \isk \partial_x \theta$, as it is sometimes done to write the bosonic fields explicitly independent of $K$, will not 
% affect the shifts of the fields, since they only depend on the commutator of $[\partial_x \theta ,\phi]$, that remains unchanged.
% At this point, we like to remind the reader, that we used the definition of fields following Ref.~\onlinecite{Gia}, that 
% sometimes differs by a sign compared to notations used by other authors. E.g. in Refs.~\onlinecite{Dolcini05,Budich}, the field $\phi$ and thus the commutator $[\phi,\partial_x \theta]$ 
% have an additional minus sign, which is why the shifts of the two fields come with opposite signs then. This will, however, coincide again for our further notation in Eq.~(\ref{eq:HR3}), since 
% then $\partial_t \phi= -vK \partial_x \theta$ with another sign. \\
Performing the shifts on the Rashba impurity in Eq.~(\ref{eq:HR}), we find with the shortcut $\gamma=\frac{i}{\pi a}\left(\frac{2 \pi a}{L}\right)^K$
% \begin{align*}
% & \Ht=\Ud \Hr U= 
% % & e^{i \hat{H}_R t} \Ud e^{-i \alpha eV i \gamma t \int dx dx' [\phi(x), \pp \partial_x \theta(x') e^{2i\phi(x')}\pp ]} U= \\
% % & e^{i \hat{H}_R t} e^{i \pi \alpha eV \gamma t  \int dx' \pp e^{2i\phi(x')}\pp }= 
% F_{+}^{\dagger}F_{-} \gamma \int dx \ \pp \left(\partial_x \theta(x) +\frac{\pi}{L}(N_{+}-N_{-}) + \frac{eV}{2 v_F} \right) e^{2i (\phi(x,t)-\frac{\pi x}{L}(N_{+}+N_{-})) }  \pp +\Hc,
% \end{align*}
% % so $\Ht=\gamma \int dx \ \pp (\partial_x \theta(x) +\frac{\omega_0}{vK} V)e^{2i\phi(x)}\pp$
% and
\begin{align}\label{eq:shiftsOnHr}
 & \Hti (t)= 
% e^{+i \hat{V}t} \gamma \int dx' \ \pp (\partial_x \theta(x') +\pi \alpha V)e^{2i\phi(x')}\pp e^{-i \hat{V}t}= \\
%  &\gamma \int dx' \ \pp (\partial_x \theta(x') +\pi \alpha V)e^{2i\phi(x')}\pp e^{+i \hat{V}t} e^{iweV 2it \int dx[\partial_x \theta(x),\phi(x')]} e^{-i \hat{V}t}=\\
%  &  \gamma \int dx' \ \pp (\partial_x \theta(x') +\pi \alpha V)e^{2i\phi(x')}\pp e^{-weV 2t \int dx (-i \pi) \delta(x-x')}= \\
 F_{+}^{\dagger}F_{-} \gamma \int dx' \ 
\pp \left(\partial_x \theta(x) +\frac{\pi}{L}(N_{+}-N_{-}) + \frac{eV}{2 v_F} \right) e^{2i (\phi(x,t)-\frac{\pi x}{L}(N_{+}+N_{-})+\frac{eV}{2} t) }  \pp +\Hc
\end{align}
% We now specify the form of the impurity and consider for simplicity a point-like Dirac-impurity of the form 
% \begin{align}\label{eq:diraclikeimpurity}
%  \alpha(x)=\alpha \delta(x).
% \end{align}
% 
% 
% Here, the Klein factors in front of the spin-orbit Hamiltonian are crucial to create the shift of the bosonic $\phi$-field.

\section{Backscattering current to second order}
\label{app:2order}
Here, we explain in detail, how to exploit normal-ordering to derive the average backscattering current in Eq.~(\ref{eq:avI1b}). \\
% A few comments are in order to explain the steps done in Eq.~(\ref{eq:avI1b}). \\
First of all, to keep the notation compact, we introduce the scalar $\nu^{-1}(\partial_t)$ with $\nu^{-1}(\partial_t)\partial_t \pp e^{2i \phi(0,t)} \pp =(vK)^{-1} \partial_t \pp e^{2i \phi(0,t)} \pp$, if $\partial_t$ acts on the bosonic part and 
$\nu^{-1}(\partial_t)  \partial_t e^{i eV t } =(v_F)^{-1} \partial_t e^{i eV t }$, if $\partial_t$ acts on the voltage part. With this, we can write Eqs.~(\ref{eq:HR3}) and ~(\ref{eq:jbsb}) in a compact form
\begin{align}
\Hti (t) &=  \frac{\alpha v_F}{2 \pi} \left(\frac{2\pi a}{L}\right)^K 
\left( \nu^{-1}(\partial_t)\ \partial_t \left( \pp e^{2i \phi(0,t)} \pp e^{i eV t }\right) +\Hc\right),\\
 \jbsti (t) &=  i e \ \frac{\alpha  v_F}{2 \pi} \left(\frac{2\pi a}{L}\right)^K 
 \left( \nu^{-1}(\partial_t)\ \partial_t \left( \pp e^{2i \phi(0,t)} \pp e^{i eV t }\right) -\Hc\right).
\end{align}
To compute $\langle \jbs \rangle$, we start from Eq.~(\ref{eq:AvFormula}),
\begin{align}\label{eq:NormalOrdering}
  \langle \jbs(t) \rangle &=  \frac{1}{2} \sum_{\eta=\pm} \left\langle T_K \left[ \jbsti^{\eta}(t) (-i) \int_K dt_2 \Hti^{\eta'}(t_2) \right] \right\rangle_0 = 
  p_1 \alpha^2 \left(\frac{2\pi a}{L}\right)^{2K} \int dt_2 \theta(t-t_2) \nu^{-1}(\partial_t) \nu^{-1}(\partial_{t_2})  \notag \\
 & \times \bigg\langle \bigg[\left( \pp \partial_t\  e^{2i   \phi(0,t)+i eV t}  \pp  - \Hc\right),
    \left( \pp \partial_{t_2}\  e^{2i   \phi(0,t_2)+i eV t_2}  \pp  + \Hc\right) \bigg] \bigg\rangle_0\notag \\
    %
% & p_1 \alpha^2 \left(\frac{2\pi a}{L}\right)^{2K} \int dt_2 \theta(t-t_2) \nu^{-1}(\partial_t) \nu^{-1}(\partial_{t_2}) \times \notag \\
% & \bigg\langle \partial_t \partial_{t_2} \bigg( h(t-t_2) [e^{i eV (t-t_2)} \pp e^{2i   \phi(0,t)} e^{-2i   \phi(0,t_2)} \pp -
%     e^{-i eV (t-t_2)} \pp e^{-2i   \phi(0,t)} e^{+2i   \phi(0,t_2)} \pp] + \notag \\
% & h(t_2-t) [e^{-i eV (t-t_2)} \pp e^{2i   \phi(0,t_2)} e^{-2i   \phi(0,t)} \pp -
%     e^{i eV (t-t_2)} \pp e^{-2i   \phi(0,t_2)} e^{2i   \phi(0,t)} \pp]
%  \bigg) \bigg\rangle_0= \notag \\
& = p_1 \alpha^2 \left(\frac{2\pi a}{L}\right)^{2K} \int dt_2 \theta(t-t_2) \nu^{-1}(\partial_t) \nu^{-1}(\partial_{t_2})  \notag \\
& \times \bigg(  \partial_t \partial_{t_2} \left( h(t-t_2) e^{i eV (t-t_2)} \right) \left\langle \pp e^{2i   \phi(0,t)} e^{-2i   \phi(0,t_2)} \pp \right\rangle_0-
    \partial_t \partial_{t_2} \left( h(t-t_2) e^{-i eV (t-t_2)} \right) \left\langle \pp e^{-2i   \phi(0,t)} e^{2i   \phi(0,t_2)} \pp \right\rangle_0 \notag \\
& + \partial_t \partial_{t_2} \left( h(t_2-t)e^{-i eV (t-t_2)} \right) \left\langle \pp e^{2i   \phi(0,t_2)} e^{-2i   \phi(0,t)} \pp \right\rangle_0-
    \partial_t \partial_{t_2} \left( h(t_2-t) e^{i eV (t-t_2)} \right) \left\langle \pp e^{-2i   \phi(0,t_2)} e^{2i   \phi(0,t)} \pp \right\rangle_0
 \bigg) \notag \\
& = (-2i) p_1 \alpha^2 \left(\frac{2\pi a}{L}\right)^{2K} \int d\tau\ \theta(\tau) \nu^{-2}(\partial_{\tau})\partial_{\tau}^2 \bigg( h(\tau) \sin(eV \tau) +h(-\tau) \sin(-eV \tau) \bigg). 
\end{align}
Here, we used $p_1= e \frac{v_F^2}{(2 \pi)^2}$
% with $p_1=\frac{1}{\hbar^3} \frac{1}{2 K} \frac{1}{(\pi a v)^2}$.
and the normal-ordering relations $\pp e^{2i \phi(t)} \pp \pp e^{-2i \phi(t_2)} \pp=\pp e^{-2i \phi(t)} \pp \pp e^{2i \phi(t_2)} \pp=h(t-t_2) \pp e^{2i \phi(t)} e^{-2i \phi(t_2)} \pp $ with
$h(t_1-t_2)=\left(2\pi/L(i v (t_1-t_2) +a) \right)^{-2K}$ \cite{Delft}. 
% The time derivatives were written outside of the normal-ordering signs, but it should be kept in mind, that operators generated by those have to remain normal-ordered. 
Importantly, time derivatives acting on the above operators will create normal-ordered operators in non-exponential form, that will again vanish under averaging, since 
per definition $\langle \pp A \pp \rangle_0=0$, but $\langle \pp e^A \pp \rangle_0=1$ for any bosonic operator $A$ \cite{Car96}. This is, however, only true, if the expectation value 
is taken with respect to the zero temperature ground state. Indeed, the expression simplifies a lot, exploiting that 
$\langle \pp e^{\pm 2i \phi(t_1)} e^{\mp 2i \phi(t_2)} \pp\rangle_0=1$. After the substitution $\tau=t-t_2$ we find the expression in the last line of Eq.~(\ref{eq:NormalOrdering}). Using the fact, that 
the functions appearing in the integrand only depend on the time difference $\tau$, we write $\partial_{t_2}=-\partial_{t}=-\partial_{\tau}$
% Moreover, since the boundaries of the time integral are 
% symmetric, we can perform a substitution $\tau \to -\tau$ at any time, taking care of arising signs. With $\theta(\tau)+\theta(-\tau)=1$, we simplify
%
to simplify
\begin{align}
 \langle \jbs \rangle & =
% p_1 \alpha^2 \left(\frac{2\pi a}{L}\right)^{2K} \int dt_2\ \theta(t-t_2) \nu^{-1}(\partial_t) \nu^{-1}(\partial_{t_2})\times \notag \\
%  & \bigg\langle \bigg[\left( \pp \partial_t\  e^{2i   \phi(0,t)+ i eV t}  \pp  - \Hc\right),
%     \left( \pp \partial_{t_2}\  e^{2i   \phi(0,t_2)+i eV t_2}  \pp  + \Hc\right) \bigg] \bigg\rangle_0 =\notag \\
% & 2i p_1 \alpha^2 \left(\frac{2\pi a}{L}\right)^{2K} \int d\tau\ \theta(\tau) \nu^{-2}(\partial_{\tau})\partial_{\tau}^2 \bigg( h(\tau) \sin(eV \tau) +h(-\tau) \sin(-eV \tau) \bigg) = \notag \\
 (-2i) p_1 \alpha^2 \left(\frac{2\pi a}{L}\right)^{2K} \int d\tau  \nu^{-2}(\partial_{\tau}) \partial_{\tau}^2 \bigg( h(\tau) \sin( eV \tau) \bigg) \notag \\
& =2i p_1 \alpha^2 \left(\frac{2\pi a}{L}\right)^{2K} \frac{(Kv -v_F)^2}{(Kv\ v_F)^2} (eV)^2 \int d\tau   h(\tau) \sin( eV \tau).
% = \notag \\
% & \frac{-2 \pi p_1 \alpha^2}{\Gamma (2 K)} \left(\frac{a}{v}\right)^{2K} \frac{(Kv -v_F)^2}{(Kv\ v_F)^2}  (eV)^{2 K+1}= \\
% & \frac{e \alpha^2}{\pi a^2 \Gamma (2 K)} \left(\frac{a}{v}\right)^{2K} \frac{(Kv -v_F)^2}{(Kv\ v_F)^2}  (eV)^{2 K+1}.
\end{align}
In the last step we applied the definition of the $\nu^{-1}(\partial_{\tau})$ and performed an integration by parts. Evaluating the integral we arrive at Eq.~(\ref{eq:avI1b}).

% \begin{align}\label{eq:avI1c}
% & \langle \jbs \rangle=
%  p_1 \alpha^2 \left(\frac{2\pi a}{L}\right)^{2K} \int d\tau \theta(\tau) 2i \partial_{\tau}^2 \bigg( h(\tau) (-\sin(2\omega_0 eV \tau)) +h(-\tau) \sin(2\omega_0 eV \tau) \bigg) = \notag \\
% & (-2i) p_1 \alpha^2 \left(\frac{2\pi a}{L}\right)^{2K} \int d\tau  \partial_{\tau}^2 \bigg( h(\tau) \sin(2\omega_0 eV \tau) \bigg)= \notag \\
% & (-2i) p_1 \alpha^2 \left(\frac{2\pi a}{L}\right)^{2K} \int d\tau  \left( \partial_{\tau}^2 h(\tau) \sin(2\omega_0 eV \tau) +2 \partial_{\tau} h(\tau) \partial_{\tau} \sin(2\omega_0 eV \tau) 
% + h(\tau) \partial_{\tau}^2 \sin(2\omega_0 eV \tau) \right)= 0. 
% \end{align}
% The last line cancels to zero, as can be seen after integrating by parts, assuming that boundary terms vanish as $h(\tau)$ decays for $K>0$.
%

\section{Fermionic approach}
\label{app:Fermionic}

\begin{figure}
\includegraphics[width=0.45 \textwidth]{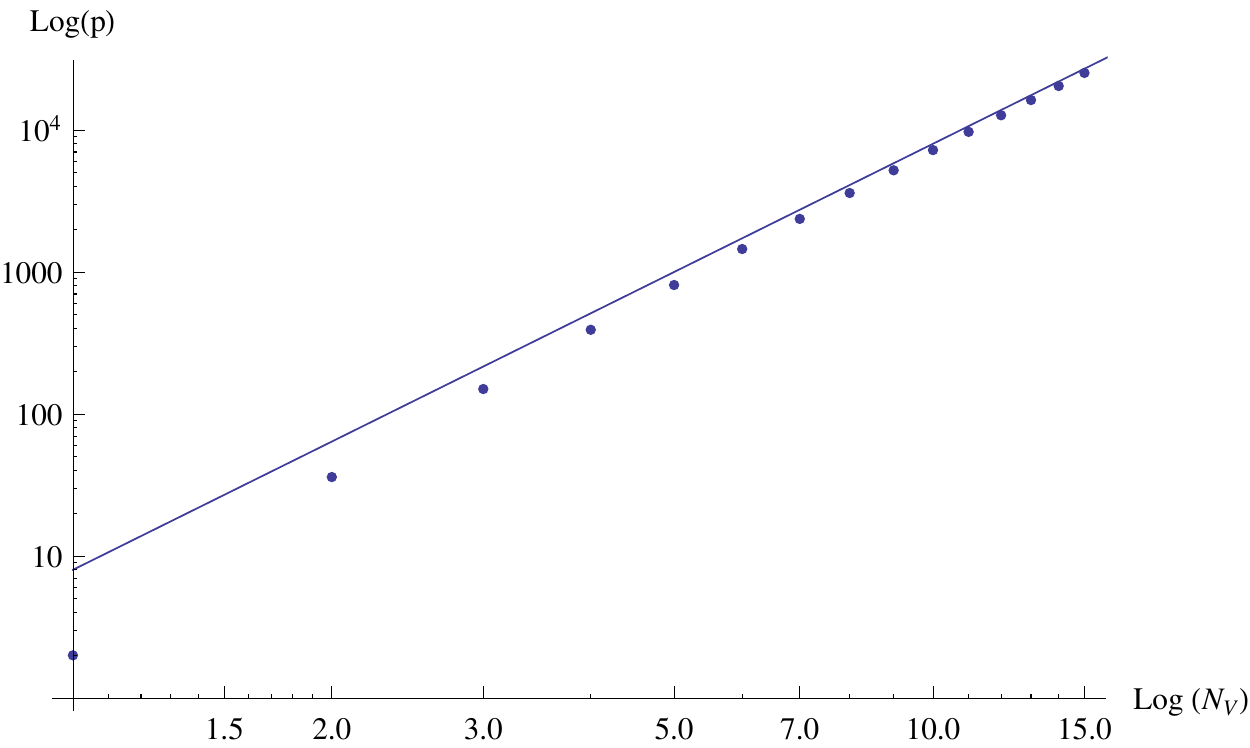}
\caption{Log-Log Plot of the numerical factor $p$ and the voltage integer $N_V$ ($N_V=\frac{L}{v_F}\frac{eV}{4\pi}$). The continuous line represents $p=\left(2 N_V\right)^3$ for comparison.}
\label{fig:VpPlot}
\end{figure}

For practical reasons, we first rewrite $T$ in Eq.~(\ref{eq:Teffective}) with positive momenta. Considering a system with only positive energies, we can transform $k\to -k$ for all left-moving states, for convenience. 
%
% The momenta are assumed to be discretized $k_i=\frac{2\pi}{L}n_i$ with integers $n_i=1,2,\ldots$ and 
The states are then labeled by their discretized momenta $k_i=\frac{2\pi}{L}n_i$, where the integers $n_i=1,2,\ldots 2 n_{FS} $ are limited by the parameter $n_{FS}$, and we find now for 
both species $\xi_{}(k_i)= v_F k_i= v_F \frac{2\pi}{L}n_i$. 
The energy of the multi-particle state is the sum of the energies of all contained single-particle states $\Ps{\pm}{k_i}$.
% A single-particle wave function we denote by $\psi_i$, where $i=\eta_i, k_i$ is a multi-index determining the species, $\eta_i=\pm$ for right and left movers, and the discretized momentum $k_i$. 
We now define the initial state $|i\rangle$ as follows: 
A Fermi sea of finite depth is represented by $n_{FS}$ occupied states on both branches. 
Next, we apply a voltage bias in such a way that the $\pm$-branches in the presence of the bias are filled up to $n_i\leq n_{FS}\pm N_{V}=n_{FS} \pm \frac{L}{v_F}\frac{eV}{4\pi}$, with $N_{V}\leq n_{FS}$.
Here, by introducing the integer $N_V$, we made the choice of applying the bias symmetrically on both branches. We thus have
\begin{align*}
|i\rangle= \left(\prod_{j=1}^{n_{FS}+N_{V}}\Psd{+}{k_j}\right)\left( \prod_{j=1}^{n_{FS}-N_{V}}\Psd{-}{k_j} \right)|0\rangle,
\end{align*}
where $|0\rangle$ is the vacuum state. Hereby, we defined a (arbitrary) reference order of the right and left moving single particle operators. This order has to be kept throughout to keep track of all signs.
The two parameters to be set in this model are $n_{FS}$ and $V$, where $n_{FS}$ can be considered as a symmetric momentum cutoff.
%
% At zero temperature, the occupation probability is $n(\epsilon_i)=1-\theta(n_i-(n_{FS}+\delta_{\eta_i,+}N_{+} + \delta_{\eta_i,-}N_{-}))$. On the other hand, 
% the chance for a final spots to be free is $\bar{n}(\epsilon_{f})= 1-n(\epsilon_{f})+\delta_{f,i}+\delta_{f,j}$, and the same for $g$.
% This is important, because final and initial state can have some overlap (in 
% contrast to a pure two-particle backscattering \cite{Crepin1}). 
%
% For the processes we consider (see Eq.~(\ref{eq:Teffective})), only one particle is backscattered, and the other particle remains on the same branch. 
%
For a general set of $n_{FS}$ and $V$, we note the following. First, as explained in the main text, the matrix element $M_{if}=0$ if all values of $q$ are taken into account. 
The cancellation mechanism is, that  one term with nonzero $q$ will always be compensated by another one with $q=0$ in any of the terms given in Eq.~(\ref{eq:Teffective}). 
If we exclude the components $q=0$ from Eq.~(\ref{eq:Teffective}) though, we get $M_{if}\neq 0$. Second, 
for a positive bias, the processes backscattering a left mover into an additional right mover, that are represented by the $\{ + \leftrightarrow - \}$ in Eq.~(\ref{eq:Teffective}), are always zero (and the other way round).
Third, the effective cutoff (or depth of the Fermi sea taken into account) $n_{FS}$ does not matter. Because of various cancellations, only processes close to the Fermi surface contribute to the backscattering current.\\ 
Adopting the same scheme as in the bosonic part, with the zero-modes decoupled from the electron interactions, we exclude the elastic $q=0$ components from Eq.~(\ref{eq:Teffective}), and subsequently employ Eq.~(\ref{eq:Ibs}) to derive the backscattering current.  
% Fixing the initial state by a choice of the parameters $n_{FS}$ and $V$, we sum over all final configurations that can appear. 
Scattering processes leading to the same final configuration will be added with their respective signs, that arise by sorting the single particle operators of 
the many particle state in the reference order. 
Adding all contributions according to Eq.~(\ref{eq:Ibs}) yields a current
\begin{align*}
 \langle \jbs \rangle=\frac{e L}{v_F} \left|\frac{-i\alpha g_2}{L}\right|^2 \ p= \frac{e \alpha^2 g_2^2}{v_F L} p.
\end{align*}
Here, we have introduced the numerical factor $p$, representing the sum of all diagrams. In general, $p$ will increase with increasing bias.
Realizing that the result does not change with the momentum cutoff, the only parameter of the system is just $V$. \\
To estimate the dependency of $p$ on the voltage $V$, we can fit a power law to the numerical data, where we would expect from the bosonic calculation (see Eq.~(\ref{eq:avI1b})) $\jbs\sim V^{2K+1}\sim V^3$ for weak interactions. 
Indeed we see that a scaling of $p \sim (2N_V)^3=(2\frac{L}{v_F}\frac{eV}{4\pi})^3$ is a reasonable approximation, particularly for large voltages (see Fig.~(\ref{fig:VpPlot})). Deviations from 
this scaling for small values of $V$ are probably due to finite size effects of the calculation. We finally arrive at Eq.~(\ref{eq:IbsFinalMain}) of the main text
\begin{align}\label{eq:IbsFinal}
 \langle \jbs \rangle \sim e \alpha^2 g_2^2 \frac{L^2}{v_F^4(2\pi)^3} (eV)^3.
\end{align}

\section{Backscattering current to fourth order}
\label{app:4order}

In this section, we give details of the calculation in fourth order of the impurity strength. This means, we expand Eq.~(\ref{eq:AvFormula}) to third order in the impurity Hamiltonian
\begin{align}\label{eq:avI4ord1}
 \langle \jbs(t_2) \rangle &= \frac{1}{2} \sum_{\eta=\pm} \left\langle T_K \left[ \jbsti^{\eta}(t_2) \frac{i}{6} \int_K dt_3 dt_4 dt_5 \Hti^{\eta'}(t_3)\Hti^{\eta''}(t_4) \Hti^{\eta'''}(t_5) \right] \right\rangle_0 \notag \\
%
% & \frac{1}{2}  \frac{1}{\hbar^2 i \pi a} \left(\frac{-1}{2vK \pi a}\right)^3 \frac{i}{6 \hbar^3} \left(\frac{2\pi a}{L}\right)^{4K} \alpha^4 \sum_{\eta, \eta', \eta'', \eta'''=\pm} \bigg\langle T_K  \intinf dt_3 dt_4 dt_5 \eta' \eta'' \eta''' \times \notag \\
% & \bigg[ 
% \bigg( \pp \ptr{2}  e^{-i2 \phin(2)-2i \omega_0 eV t_2} \pp -\Hc \bigg)  \bigg( \pp \ptr{3}  e^{-i2 \phinp (3)-2i \omega_0 eV t_3} \pp +\Hc \bigg)\times \notag\\
% & \bigg( \pp \ptr{4}  e^{-i2 \phinpp(4)-2i \omega_0 eV t_4} \pp +\Hc \bigg) \bigg( \pp \ptr{5}  e^{-i2 \phinppp(5)-2i \omega_0 eV t_5} \pp +\Hc \bigg)
% \bigg] \bigg\rangle_0=\notag\\
%
% & \frac{1}{2}  \frac{e}{2 \pi^2 \hbar a}\frac{i}{6 \hbar^3} \frac{i^3}{(\pi a)^3}\left(\frac{2\pi a}{L}\right)^{4K} \sum_{\eta, \eta', \eta'', \eta'''=\pm}  {2\pi} \int dx_2 dx_3 dx_4 dx_5 \intinf dt_3 dt_4 dt_5 \eta' \eta'' \eta''' n^4\times \notag \\
% & \langle T_K \bigg[ 
% \sum_{\lambda_3, \lambda_4, \lambda_5=\pm }(C^{(-) \lambda_3 \lambda_4 \lambda_5}_{\eta,\eta',\eta'',\eta'''}(2,3,4,5)-C^{(+) \lambda_3 \lambda_4 \lambda_5}_{\eta,\eta',\eta'',\eta'''}(2,3,4,5)
% \bigg] \rangle_0=  \notag\\
& =\frac{p_2}{12} \left(\frac{2\pi a}{L}\right)^{4K} \alpha^4 \sum_{\eta, \eta', \eta'', \eta'''=\pm}  \intinf dt_3 dt_4 dt_5 \eta' \eta'' \eta'''  \notag \\
& \times \bigg\langle T_K 
\sum_{\lambda_3, \lambda_4, \lambda_5=\pm }\left(C^{(+) \lambda_3 \lambda_4 \lambda_5}_{\eta,\eta',\eta'',\eta'''}(2,3,4,5)- \{\lambda_i\to -\lambda_i \}_{i \in \{2,3,4,5\}} \right) \bigg\rangle_0.
\end{align}
With $p_2=
% -\frac{1}{2}  \frac{1}{\hbar^2 i \pi a v} \left(\frac{-1}{2vK \pi a}\right)^3 \frac{i}{6 \hbar^3}=
% \frac{1}{(2\pi a v)^4 K^3}$.
\frac{-e v_F^4}{(2\pi)^4}$.
Here we have defined a shorthand notation
\begin{align}\label{eq:avI4ord2}
 \ellll & \equiv  
% \bigg( \pp \ptr{2}  e^{-i2 \lambda_2 \phin(2)-2i \omega_0 eV t_2} \pp -\Hc \bigg)  \bigg( \pp \ptr{3}  e^{-i2 \lambda_3 \phinp (3)-2i \omega_0 eV t_3} \pp +\Hc \bigg)\times \notag\\
% & \bigg( \pp \ptr{4}  e^{-i2 \lambda_4 \phinpp(4)-2i \omega_0 eV t_4} \pp +\Hc \bigg) \bigg( \pp \ptr{5}  e^{-i2 \lambda_5 \phinppp(5)-2i \omega_0 eV t_5} \pp +\Hc \bigg)
% \bigg]
%  e^{-2i\omega_0 eV (\lambda_2 t_2 +\lambda_3 t_3+ \lambda_4 t_4 + \lambda_5 t_5)}=\notag \\
%
% &
 \nu^{-1}(\partial_{t_2}) \nu^{-1}(\partial_{t_3}) \nu^{-1}(\partial_{t_4}) \nu^{-1}(\partial_{t_5}) \notag \\
& \times \bigg(\ptr{2} \ptr{3} \ptr{4} \ptr{5} \bigg[ 
\left(h(2,3)h(2,4)h(2,5) h(3,4) h(3,5) h(4,5) \right) e^{i eV (\lambda_2 t_2 +\lambda_3 t_3+ \lambda_4 t_4 + \lambda_5 t_5)} \bigg] \bigg)  \notag \\
&\times  \pp e^{i\lambda_2 2 \phin(2)}  e^{i\lambda_3 2 \phinp(3)} e^{i\lambda_4 2\phinpp(4)} e^{i\lambda_5 2 \phinppp(5)} \pp
\end{align}
$\lambda_i=\pm 1$ just represents a sign here and $h(t_i,t_j)=\left(2\pi/L(i v (t_i-t_j) +a) \right)^{2K \lambda_i \lambda_j}$ with $i,j \in \{2,3,4,5\}$\cite{Delft}. 
% Obviously, the sign of the power depends on which operators are normal-ordered together. 
The neutrality rule \cite{Gia} holds overall, such $\sum_{i=2}^{5} \lambda_i=0$. From the same arguments used in the previous section, we argue, that the derivations act 
only on the terms in front of the operators in normal ordering signs at zero temperature.\\
We can see from Eq.~(\ref{eq:avI4ord2}), that $\ellll $ stays invariant under a simultaneous change 
of two times $t_i$ and $t_j$, $\lambda_i$ and $\lambda_j$ (and the Keldysh index $\eta_i$ and $\eta_j$). Importantly, this is only true if the order of operators $e^{i\lambda_i \phi(i)}$ can be interchanged, 
which is possible before applying $T_K$, \textit{but not after}. From this fact we infer, that we can change for instance $\epmpm$ and $\epmmp$ to $\eppmm$ in Eq.~(\ref{eq:avI4ord1}), since the time variables $t_3,t_4,t_5$ are integrated over and 
can just be renamed (this brings a factor of $3$ in front). We get
\begin{align}\label{eq:avI4ord4}
 \langle \jbs(t_2) \rangle &=
%  -\alpha \left(\frac{2\pi a}{L}\right)^{4K} \sum_{\eta, \eta', \eta'', \eta'''=\pm} \int dx_2 dx_3 dx_4 dx_5 \intinf dt_3 dt_4 dt_5 \eta' \eta'' \eta''' n^4\times \notag \\
% & \langle T_K 3\bigg[ 
% (\eppmm-\Hc)- (\emmpp-\Hc)
% \bigg] \rangle_0=  \notag \\
 \frac{p_2}{4} \alpha^4 \left(\frac{2\pi a}{L}\right)^{4K} \sum_{\eta, \eta', \eta'', \eta'''=\pm}  \intinf dt_3 dt_4 dt_5 \eta' \eta'' \eta'''  \notag \\
& \times \left\langle T_K \left(\eppmm-\emmpp\right) \right\rangle_0.
\end{align}
% In the last step we used that $C^{--++}$ is exactly the hermitian conjugate of $C^{++--}$.

The next step will be explicit time-ordering.
Because multi-dimensional, coupled integrals are involved, we were not able to find an analytic result for the average backscattering current. However, to proceed, 
two different approximations are studied, in the form of contractions in time. Physically, we assume, that among the four operators with four different times occuring in the integrand, some
operators couple strongly to each other, while others only couple weakly. 
This procedure is mathematically justified under certain conditions, that we specify below.
\\

\paragraph*{Time-ordering}
% For now, we look for the terms in lowest order of $V$. Therefore we take 
% \begin{align}\label{eq:DefExp}
% % & e^{\lambda_2 \lambda_3 \lambda_4 \lambda_5}(2,3,4,5)-\Hc= 
% % \bigg[\bigg( \frac{1}{2^4 K^4 v^4} \ptr{2} \ptr{3} \ptr{4} \ptr{5} \bigg)
% % \left(h(2,3)h(2,4)h(2,5) h(3,4) h(3,5) h(4,5) \right)\bigg]\times \notag \\
% % &  \pp e^{i\lambda_2 \phi(2)}  e^{i\lambda_3 \phi(3)} e^{i\lambda_4 \phi(4)} e^{i\lambda_5\phi(5)} \pp  
% % e^{2iv\tilde{V} K (\lambda_2 t_2 +\lambda_3 t_3+ \lambda_4 t_4 + \lambda_5 t_5)}-\Hc +\mathcal{O}(\tilde{V}).
% & \eppmm-\Hc= 
% \bigg[\bigg( \frac{1}{2^4 K^4 v^4} \ptr{2} \ptr{3} \ptr{4} \ptr{5} \bigg)
% \left(h(2,3)h(2,4)h(2,5) h(3,4) h(3,5) h(4,5) \right) \notag \\
% & e^{-2iv\tilde{V} K ( t_2 + t_3- t_4 - t_5)}\bigg]\times  \pp e^{i 2\sk \phin(2)}  e^{i2\sk \phinp(3)} e^{-i2\sk \phinpp(4)} e^{-i 2\sk\phinppp(5)} \pp  
% -\Hc
% \end{align}
% This term is obviously the lowest order term, as long as the time-derivatives do not translate into additional factors of $V$! \\
% Importantly, from the fixed signs of the exponents, we note that $h(2,3), h(3,2)$ and $h(4,5),h(5,4)$ always have exponents $+2K$, while all the other functions $h$ have exponents $-2K$.
We continue from Eq.~(\ref{eq:avI4ord4}) and split the full expression into two parts where the one depends explicitly on the Keldysh indizes, and the other does not.
\begin{align}\label{eq:avI4ord5}
& \langle \jbs(t_2) \rangle=  
% 3p_2 \alpha^4 \left(\frac{2\pi a}{L}\right)^{4K} \sum_{\eta, \eta', \eta'', \eta'''=\pm} \intinf dt_3 dt_4 dt_5 \eta' \eta'' \eta''' \times  \notag \\
% & \bigg( \bigg\langle T_K \bigg( \ptr{2} \ptr{3} \ptr{4} \ptr{5} \bigg)\bigg[
% \left(h(2,3)h(2,4)h(2,5) h(3,4) h(3,5) h(4,5) \right) e^{-2iv\tilde{V} K ( t_2 + t_3- t_4 - t_5)}\bigg]\times \notag \\
% &  \pp e^{i 2\sk \phin(2)}  e^{i2\sk \phinp(3)} e^{-i2\sk \phinpp(4)} e^{-i 2\sk\phinppp(5)} \pp  
% \bigg\rangle - \notag \\
% &\bigg\langle T_K \bigg(  \ptr{2} \ptr{3} \ptr{4} \ptr{5} \bigg)\bigg[
% \left(h(2,3)h(2,4)h(2,5) h(3,4) h(3,5) h(4,5) \right) e^{+2iv\tilde{V} K ( t_2 + t_3- t_4 - t_5)}\bigg]\times \notag \\
% &  \pp e^{-i 2\sk \phin(2)}  e^{-i2\sk \phinp(3)} e^{+i2\sk \phinpp(4)} e^{+i 2\sk\phinppp(5)} \pp  
% \bigg\rangle\bigg) =\notag \\
%
% &  3p_2 \alpha^4 \left(\frac{2\pi a}{L}\right)^{4K} \frac{1}{2^4 K^4 v^4} \sum_{\eta, \eta', \eta'', \eta'''=\pm} \intinf dt_3 dt_4 dt_5 \eta' \eta'' \eta''' \times  \notag \\
% & \bigg\langle T_K  \pp e^{i 2\sk \phin(2)}\pp \pp e^{i2\sk \phinp(3)} \pp \pp e^{-i2\sk \phinpp(4)}\pp \pp e^{-i 2\sk\phinppp(5)} \pp  \bigg\rangle \times \notag \\
% & \left(\frac{\bigg(  \ptr{2} \ptr{3} \ptr{4} \ptr{5} \bigg)\bigg[
% \left(h(2,3)h(2,4)h(2,5) h(3,4) h(3,5) h(4,5) \right) e^{-2iv\tilde{V} K ( t_2 + t_3- t_4 - t_5)}\bigg]}{\left(h(2,3)h(2,4)h(2,5) h(3,4) h(3,5) h(4,5) \right)}\right)\bigg|_{ord} -\Hc\equiv \notag \\
 \frac{p_2}{4} \alpha^4 \left(\frac{2\pi a}{L}\right)^{4K} \intinf dt_3 dt_4 dt_5 \ \cnull \left( \ceinsp -\ceinsm \right).
\end{align}
We undid the full normal-ordering to apply Wick's theorem in the following.
Now, the terms are
\begin{align*}
 \cnull&=\sum_{\eta, \eta', \eta'', \eta'''=\pm}  \eta' \eta'' \eta''' \bigg\langle T_K  \pp e^{i 2  \phin(2)}\pp \pp e^{i2  \phinp(3)} \pp \pp e^{-i2  \phinpp(4)}\pp \pp e^{-i 2 \phinppp(5)} \pp  \bigg\rangle_0 \notag \\
& =\sum_{\eta, \eta', \eta'', \eta'''=\pm}  \eta' \eta'' \eta''' \bigg\langle T_K  \pp e^{-i 2  \phin(2)}\pp \pp e^{-i2  \phinp(3)} \pp \pp e^{+i2  \phinpp(4)}\pp \pp e^{+i 2 \phinppp(5)} \pp  \bigg\rangle_0, \\
\ceinspm &= \nu^{-1}(\partial_{t_2}) \nu^{-1}(\partial_{t_3}) \nu^{-1}(\partial_{t_4}) \nu^{-1}(\partial_{t_5}) \notag \\
& \times \left(\frac{\bigg(  \ptr{2} \ptr{3} \ptr{4} \ptr{5} \bigg)\bigg[
\left(h(2,3)h(2,4)h(2,5) h(3,4) h(3,5) h(4,5) \right) e^{\pm i eV ( t_2 + t_3- t_4 - t_5)}\bigg]}{\left(h(2,3)h(2,4)h(2,5) h(3,4) h(3,5) h(4,5) \right)}\right)\bigg|_{ord}.
\end{align*}
$\cnull$ is the same for both configurations, because of the Debye-Waller formula \cite{Gia}. 
% We see from $C_1$, that the operation $\lambda_i \to -\lambda_i$ can \textit{not} 
% be identified with a hermitian conjugation, since the functions $h$ are not real.
% While $C_0$ includes all the field operators and thus explicitly depends on the Keldysh indizes, the functions $C_1$ do not, but incorporate time-ordering in a subtle way.
Since the average acts only on operators, we can pull out the factor including the time derivatives, keeping in mind, that this factor depends on time ordering, indicated by the label $|_{ord}$.
% Obviously, the factors $C_1$ follows the time-ordering of the exponential operators. 
However, it is still a scalar, which 
is why we can use Wick's theorem to simplify the product $C_0$ of the exponential operators. \\
% After normal ordering, the term $C_0$ will always be of a form to cancel the denominator in the factors $C_1$, 
% and we will be left with the time-ordered derivatives in the nominator of $C_1$.\\
The next step is to perform time-ordering explicitly, so to evaluate the sum over Keldysh indizes.
% \begin{align*}
%  X\equiv \sum_{\eta, \eta', \eta'', \eta'''=\pm} \eta' \eta'' \eta''' \bigg\langle T_K  \pp e^{i 2  \phin(2)}\pp \pp e^{i2  \phinp(3)} \pp \pp e^{-i2  \phinpp(4)}\pp \pp e^{-i 2 \phinppp(5)} \pp  \bigg\rangle.
% \end{align*}
Hereby, we use Wick's theorem, which for exponential operators is nothing but the Debye-Waller relation
\begin{align}\label{eq:toExpl}
 \cnull & = 
\sum_{\eta, \eta', \eta'', \eta'''=\pm}\ \eta' \eta'' \eta''' \prod_{i< j, (i,j) \in (2,3,4,5)}
 \bigg\langle T_K  \pp e^{i 2  \lambda_i \phi^{\eta_i}(x)}\pp \pp e^{i2  \lambda_j \phi^{\eta_j}(y)} \pp \bigg\rangle_0 \notag \\
& =\sum_{\eta_2, \eta_3, \eta_4, \eta_5=\pm} \eta_3 \eta_4 \eta_5\ C_0^{\eta_2 \eta_3}(2,3) C_0^{\eta_2 \eta_4}(2,4) C_0^{\eta_2 \eta_5}(2,5) C_0^{\eta_3 \eta_4}(3,4) C_0^{\eta_3 \eta_5}(3,5) C_0^{\eta_4 \eta_5}(4,5).
\end{align}
For now, $\eta_i, \eta_j \in (\eta_2, \eta_3, \eta_4, \eta_5)$ have to be identified with the $(\eta, \eta', \eta'', \eta''')$.
We defined\\
$C_0^{\eta_i \eta_j}(x,y)= \langle T_K  \pp e^{i 2  \lambda_i \phi^{\eta_i}(x)}\pp  \pp e^{i2  \lambda_j \phi^{\eta_j}(y)} \pp \rangle_0$. At this point, 
we have to explicitly perform the time-ordering. To do so, we specify:
\begin{align*}
 & C_0^{++}(x,y)=h(x,y)\theta(x-y)+h(y,x)\theta(y-x), \\
& C_0^{+-}(x,y)=h(y,x), \\
& C_0^{-+}(x,y)=h(x,y), \\
& C_0^{--}(x,y)=h(x,y) \theta(y-x)+h(y,x) \theta(x-y).
\end{align*}
With this, we perform the sum in Eq.~(\ref{eq:toExpl}), eliminating along the way mute combinations with unphysical, cyclic time-ordering.\\

% We now work out all the contributions in term of operator arrangements. With this I mean,that we consider for example first all terms where the operator $e^{\phi(t_2)}$ is left of
% $e^{\phi(t_3)}$, which is left of $e^{-\phi(t_4)}$ and $e^{-\phi(t_5)}$. This arrangement let us call for example $\{t_2,t_3,t_4,t_5\}$. In total there are $24$ combinations like this. 
% The configuration $\{t_2,t_3,t_4,t_5\}$ in general does not mean that $t_2>t_3>t_4>t_5$ (only if all of the operators are on the $+$-branch), but it can arise
% from different time-ordering configurations. We find for example that for this case we have
% \begin{align}\label{eq:2345noContr}
% & \cnull= h(2,3)h(2,4)h(2,5) h(3,4) h(3,5) h(4,5)\times \notag \\
% &\bigg( -\theta(3>2)\theta(4>5) + \theta(3>4>5) + \theta(4>3>2) + \theta(2>3>4>5) - \theta(5>4>3>2) \bigg).
% \end{align}
% Those combinations come from configurations where the operators are distributed differently on the $+$ and $-$ Keldysh branch, but all result in the operator order 
% $e^{\phi(t_2)+\phi(t_3)-\phi(t_4)-\phi(t_5)}$. The expression for $\ceins$ is too long to be given here. Instead we will directly contract those terms. \\
% 
% % To simplify, we use contractions to focus on two-particle backscattering.

\paragraph*{TPB Contractions}

% \subsection{TPB $e^{4\phi}e^{-4\phi}$}
In this subsection, we contract times with a focus on the possibility of TPB. Those are processes, where 
two particles are backscattered from left-movers to right-movers in a very short sequence, and 
the reversed process takes places at another time. In the contraction process, we use the abbreviations $y=t_2-t_3$, $y'=t_4-t_5$, $Y=(t_2+t_3)/2$, $Y'=(t_4+t_5)/2$.
Then, we replace the integration variables $t_2, t_3, t_4$ by $y,y',Y'$. The contraction implies, that $y,y'<<Y,Y'$, so we neglect the $y,y'$ whereever safely possible, treating the cutoff 
$a$ with care. \\
We now Taylor expand the expression around the small times $y,y'\sim 0$,
% 
%  Before doing so, we set $a\sim 0$, since $a$ is very small as well and we do not want to neglect $y,y'$ against $a$, but only against $Y,Y'$. The expansion is thus done at $a=0$ to cure this problem, and the 
% cutoff is put back afterwards.\\
and find terms of different powers of the times $y,y'$ and $(Y-Y')$. In agreement with the spirit of the contraction, we focus on leading powers, that are $2K-2$ for the small times $y,y'$, and $-8K$ for the time $(Y-Y')$. 
From these powers we can infer, that the TPB contraction is only well-justified for $K<1/2$, since only then the functions containing the small time variables $y,y'$ decay fast enough. On the other hand, 
for $K>1/2$, the TPB contributions selected by this contraction procedure, still exist, but will be dominated by other terms where the $y,y'$ are not small time distances any more.
% Besides, for $K>1/2$, the functions $y^{2K-2}$ are the only ones to decay for large times, while 
% the next higher-order terms $y^{2K}$ are not. This means that a TPB contraction is in general not justified for those terms. 

Proceeding as described above, we find in lowest order the full term

\begin{align}\label{eq:C0andC1}
& \cnull \left(\ceinsp-\ceinsm \right)  \notag \\
& =\frac{1}{(vK)^4}\left(\frac{2\pi}{L}\right)^{-4K} 
(-16) (1-2 K)^2 K^2 v^4 (s_1-s_2)
   (s_1 (L_2-L_3)+s_2
   (L_1-L_3)).
\end{align}
Eq.~(\ref{eq:C0andC1}) can now be applied to Eq.~(\ref{eq:avI4ord5}), after evaluating the three different types of integrals: 
\begin{align}\label{eq:largeY}
&  L_{1/2}\sim \int  dY (\mp iv Y+a)^{-8K} \theta(Y) \left(e^{-2i eV Y}-e^{+2i eV Y}\right) =  -i 2^{8K} e^{\pm 4i K \pi}\cos(4 K \pi) \Gamma(1 - 8 K) (eV/v)^{8K-1} v^{-1}, \\
% &  \int  dY (+iv Y+a)^{-8K} \theta(Y) \left(e^{-4i \omega_0 eV Y}-e^{+4i  \omega_0 eV Y}\right) \sim -\frac{1}{2} i e^{-4 i \pi  K} v^{-8 K} \cos (4)(\omega_0 V)^{8K-1}, \\
&  L_3 \sim \int dY ((v Y)^2)^{-4K} \theta(Y) \left(e^{-2i eV Y}-e^{+2i eV Y}\right) = -i 2^{8K} \cos(4 K \pi) \Gamma(1 - 8 K) (eV/v)^{8K-1} v^{-1}, 
% &  Lon_{1/2}=\int  dY (\mp iv Y+a)^{-8K} \theta(Y) \left(e^{-2i eV Y}+e^{+2i eV Y}\right) =  2^{8K} e^{\pm 4i K \pi}\sin(4 K \pi) \Gamma(1 - 8 K) (eV/v)^{8K-1} v^{-1}, \\
% % &  \int  dY (+iv Y+a)^{-8K} \theta(Y) \left(e^{-4i \omega_0 eV Y}-e^{+4i  \omega_0 eV Y}\right) \sim -\frac{1}{2} i e^{-4 i \pi  K} v^{-8 K} \cos (4)(\omega_0 V)^{8K-1}, \\
% &  Lon_3=\int dY ((v Y)^2)^{-4K} \theta(Y) \left(e^{-2i eV Y}+e^{+2i eV Y}\right) =  2^{8K} \sin(4 K \pi) \Gamma(1 - 8 K) (eV/v)^{8K-1} v^{-1}.
\end{align}
Hereby, we substituted for simplicity $Y-Y'\to Y$. The short time integrals read
\begin{align}\label{eq:smally}
 & s_{1/2} \sim \int_{-x}^{x}  dy (\pm ivy+a)^{+2K-2} \theta(y)=\mp \frac{i a^{2 K-1}}{(1-2 K) v}+\frac{e^{\pm i \pi K} (v x)^{2 K-1}}{(1-2 K) v}.
\end{align}
%s_1=i1a, s_2=i1b
%
% We can see, that the last, big bracket contains only sums of imaginary counterparts, so it corresponds to a real expression. In contrast, the first two brackets 
% % $\left((-i v y+a)^{2 K-2}-(i v y+a)^{2 K-2}\right)$ 
% represent purely imaginary expressions. This means, that the overall expression is real, as it should be.\\

In Eq.~(\ref{eq:smally}), the integration range was limited by a general threshold $x$, which is introduced, since in the contraction process, the times $y,y'$ are assumed to be small. Moreover, we approximated
$x \gg a$.
Eq.~(\ref{eq:smally}) indicates, that the times we consider as small in the contraction process are described by a sufficiently fast decaying power law only in the regime $K<1/2$, so in this case 
the TPB contraction is justified.\\
For such strong interactions, the TPB processes are of leading power in the bias and we can find from Eqs.~(\ref{eq:avI4ord5}) and (\ref{eq:C0andC1})
\begin{align}\label{eq:jbsRKsmNo2}
 \langle \jbs \rangle \sim
  \frac{\pi  2^{8 K+5} }{K^2 \Gamma (8 K)} \frac{e \alpha^4}{4 (2 \pi)^4} \left(\frac{a}{v}\right)^{8K-2} \left(\frac{v_F}{v}\right)^4 (eV)^{8 K-1}, \hspace{1cm} \text{if}\ \ K<1/2.
\end{align}
As a nice verification, the very same result can as well be achieved by integrating Eq.~(\ref{eq:avI4ord5}) 
by parts. From all the possible terms arising, we can then source out the only one of the power $V^{8K-1}$, that is exactly Eq.~(\ref{eq:jbsRKsmNo2}).\\ \\
At the point $K=1/2$, the contraction process stipulates, that the average current vanishes (see Eq.~(\ref{eq:smally})). 
For the case $K>1/2$, the TPB contractions are not well justified any more. Assuming that we still follow the same subset of terms, the $x$-dependent term in Eq.~(\ref{eq:smally}) will be of importance. We argue, that the time integration is then restricted by the next greater energy scale, being $x \to 1/eV$. By this virtue, 
we derive the scaling of the backscattering current of $\sim V^{4K+1}$. This means, that we expect a crossover of scales from $V^{8K-1}$ to $V^{4K+1}$ at the point $K=1/2$. \\

\paragraph*{SPB Contractions}

Alternatively to the contractions described above, we can also aim for SPB contributions. 
Hereby, we contract other time-variables together, that come with opposite signs in the exponents. For terms $C_1^{\pm}$, as they occur in Eq.~(\ref{eq:avI4ord5}), we 
define for instance $y=t_2-t_4$, $y'=t_3-t_5$, $Y=(t_2+t_4)/2$, $Y'=(t_3+t_5)/2$, and approximate $y,y'<< Y,Y'$. This corresponds to the case where two SPB events 
are separated in time by a large distance, and thus decouple. It turns out, that 
all the possibilities to contract in this way have to be added up to come to a meaningful result. We find 
\begin{align}\label{eq:jbsRKlaNo2}
 & \langle \jbs \rangle  \sim  \frac{(v K-v_F)^4}{(vK)^4} \frac{\alpha^4 e}{16 \pi^2 \Gamma (2 K)^2} \left(\frac{a}{v}\right)^{4K+1} (eV)^{4 K+2}.
\end{align}
%
% For the shot noise, the parts diverging at zero frequency in Eq.~(\ref{eq:Defnoise}) have to be treated with care to eventually cancel each other. 
Interestingly, this scaling of $\alpha^4\ V^{4K+2}$ in Eq.~(\ref{eq:jbsRKlaNo2}) is achieved in the following way: After the contractions, terms proportional to $(Y − Y')$
cancel altogether, while from the time-ordering we are left with some Heaviside-functions to be evaluated $ \int dY \theta(Y −Y' )\theta (Y' − Y) \sim 1/4 \frac{a}{v}$. It can be seen, that with this approximation, the correlations between the backscattering
processes are cut, leaving behind two completely independent SPB events. Those, however, always have the same ``center of mass''-times $Y=Y'$.
Because of the mentioned decoupling, one time-integral can not contribute a factor of $1/V$, increasing the scaling by one. Another option would be to contract three times together, 
which is expected to yield SPB contributions of the scaling $\alpha^4 V^{2K+1}$, adding corrections to the second order expression.

\bibliography{LuttBib4}

\end{document}